\documentclass[12pt]{article}
\pdfoutput=1 
\usepackage{jheppub}
\usepackage{graphicx,verbatim}
\usepackage{psfrag, times}
\usepackage[T1]{fontenc}

\usepackage{epsf}

\def\be{\begin{equation}}
\def\ee{\end{equation}}
\def\bc{\begin{center}}
\def\ec{\end{center}}

\def\Zinst{{\mathbb{Z}}^{\rm inst.}}

\def\cH{{\mathcal{H}}}

\def\cN{{\mathcal{N}}}

\def\cJ{{\mathcal{J}}}

\def\cY{{\mathcal{Y}}}
\def\cZ{{\mathcal{Z}}}

\def\nn{\nonumber}

\def\ep{{\epsilon}}

\def\tm{{\tilde{m}}}

\def\r2{{\sqrt{2}}}

\def\Yl{Y_{l}}
\def\Yk{Y_{k}}

\def\Nc{N_c}
\def\tM{\tilde{M}}
\def\be{\begin{equation}}
\def\ee{\end{equation}}
\def\bea{\begin{eqnarray}}
\def\eea{\end{eqnarray}}
\def\qmo{Q_{m_{1}}}
\def\qmt{Q_{m_{2}}}
\def\qfo{Q_{f_{1}}}
\def\rot{R_1^T}
\def\rtt{R_2^T}
\def\rtth{R_3^T}
\def\aro{|R_1|}
\def\art{|R_2|}
\def\arth{|R_3|}
\def\nn{\nonumber}
\def\half{{1 \over 2}}
\def\hqmo{{\hat Q}_{m_{1}}}
\def\hqmt{{\hat Q}_{m_{2}}}
\def\hqfo{{\hat Q}_{f_{1}}}

\def\hmo{{\hat m}_1}
\def\hmt{{\hat m}_2}
\def\bP{\bf{P}}

\begin{document}
\title{\Large On Integrable Structure and Geometric Transition in Supersymmetric Gauge Theories}
\author[]{Heng-Yu Chen${}^{1}$ and Annamaria Sinkovics${}^{2,3}$}
\affiliation{$^1$Department of Physics and Center for Theoretical Sciences, \\
National Taiwan University, Taipei 10617, Taiwan}
\affiliation{$^2$DAMTP, Centre for Mathematical Sciences, University of Cambridge, Wilberforce Road, Cambridge CB3 0WA, UK}
\affiliation{$^3$ Institute of Theoretical Physics, MTA-ELTE Theoretical Research Group, E{\"o}tv{\"o}s Lor{\'a}nd University, 1117 Budapest, P{\'a}zm{\'a}ny s. 1/A, Hungary}
\emailAdd{ heng.yu.chen@phys.ntu.edu.tw} \emailAdd{sinkovics@general.elte.hu}
\vspace{2cm}
\abstract
{We generalize the exact field theoretic correspondence proposed in \cite{DHL} and embed it into the context of refined topological string. The correspondence originally proposed from the common integrable structures in different field theories can  be recast as a special limit of the refined geometric transition relating open and closed topological string partition functions. We realize the simplest examples of the correspondence explicitly in terms of open-closed geometric transition. }

\maketitle

\section{Introduction}
\paragraph{}
Quantum vacua of supersymmetric gauge theories in different dimensions can sometimes share identical underlying mathematical structures, often these structures are associated with various integrable systems such as spin chains, Nahm equation or more generally Hitchin integrable system. Beyond being merely a mathematical coincidence, it is interesting to ask about the precise connection between these theories, which can lead us to uncover exciting new correspondences; moreover given many of the field theories have explicit realizations in string theory, we can also ask whether these correspondences have geometric origin. 

As an initial step, in \cite{DHL, CDHL} a new exact correspondence between the quantum vacua of four dimensional $\cN=2$ supersymmetric gauge theories and two dimensional $\cN=(2,2)$ gauged linear sigma models was proposed, as both sets of vacua can be identified with the eigenstates of the same quantum integrable Hamiltonian under specific conditions for the parameters. More explicitly, here we introduced the so-called Omega deformation in two of the four dimensions in the $\cN=2$ gauge theory  \cite{NSlimit}, and the resultant equivariant prepotential was proposed to be the Yang-Yang functional \cite{YangYang} for quantizing the classical integrable system from the Seiberg-Witten curve. Interestingly if we assign the appropriate matter contents and mass parameters to the gauged linear sigma model, the effective twisted superpotential also became the generating functional for the algebraic Bethe ansatz of the same integrable system.
It seemed that such a correspondence emerges in a rather abstract or even unintuitive way, however through a Hanany-Witten type D-brane construction \cite{Hanany1996, Hanany2004} the two dimensional gauged linear sigma model can be interpreted as the world volume theory of the co-dimension two defects or vortices in the four dimensional theory. It is this picture which allows us to explore the geometric origin of the correspondence. 

In this note we shall consider an alternative but equivalent string realization of the aforementioned field theories\footnote{More precisely, their K-theoretic generalizations.}.  Using the refined topological strings\cite{IKV, Taki2007, Awata2008} \footnote{The equivalence between Hanany-Witten type D-brane and geometric engineering constructions \cite{Katz1996, Klemm1996} of four dimensional $\cN=2$ supersymmetric gauge theories was first discussed in \cite{Karch1998}.} it turns out that the theories on  both sides of correspondence can be realized as open (2d) and closed (4d) refined topological string amplitudes respectively. From such a realization, the geometric origin of correspondence becomes manifest, that is we need to relate open and closed topological string amplitudes.  This is precisely given by a refined generalization of ``geometric transition''  proposed in \cite{Gopakumar1998},  for some interesting recent work on refined geometric transition see also  \cite{Dimofte2010, Taki2010, AS2011, AS2012}.

\section{Nekrasov-Shatashivili limit and Saddle Point Approximation}
\paragraph{}
Here we begin by discussing the simple K-theoretic generalization of the exact correspondence proposed in \cite{DHL, CDHL},  and in the later sections we shall consider the realization of these gauge theories in the  M-theory interpretation of the refined topological string, and how the exact correspondence discovered in \cite{DHL} can be recast as a beautiful realization of refined geometric transition. 
We can easily generalize the result
proved in \cite{CDHL} by attaching to them a compactified $S^1$ of radius $R$, which can be identified with the M-theory cycle. As the result, on one side of the correspondence we have:

{\bf Theory I}:  Five dimensional $\cN=1$ Supersymmetric QCD on $R^4 \times S^1$ with gauge group $U(N_c)$, with $\Nc$ fundamental hypermultiplets of masses $m_l,~ l=1,\dots \Nc$, and $\Nc$ anti-fundamental hypermultiplets of masses $\tm_l,~ l=1,\dots, \Nc$. This theory carries a complex coupling $\tau=\frac{4\pi i}{g^2} +\frac{\vartheta}{2\pi}$.

Theory I here is subjected to a special limit of so-called Omega deformation in $R^4$ proposed by Nekrasov and Shatashvili \cite{NSlimit}, given by the following twisted boundary condition $(\vec{x}, x^5)\sim (\exp(R\Omega) \vec{x}, x^5+2\pi R)$ with $\Omega={\rm diag}(\epsilon {\bf 1}_2, {\bf 0}_2)$.
We shall hereafter refer to this limit as ``NS-limit'' or "NS-deformation", this preserves three dimensional $\cN=2$ supersymmetry in $R^2\times S^1 \subset R^4\times S^1$. The NS deformation can be interpreted as turning on quantized electromagnetic flux lines in $R^4$,  as a result the continuous Coulomb branch moduli space is now lifted to only a set of discrete points  given by:
\begin{equation}\label{HiggsRoot}
a_l=m_l-n_l\epsilon,  \quad l=1,\dots, \Nc.
\end{equation}
Here $\{a_l\}$ are the vevs of the adjoint scalar in the vector multiplet, and $n_l \in {\mathbb Z}$ is the unit of quantized electromagnetic flux under $l$-th $U(1)$ factor, and as discussed extensively in \cite{DHL}, the condition (\ref{HiggsRoot}) can be interpreted as the quantization of a special locus in the moduli space of Theory I. In particular when $n_l=0$ (\ref{HiggsRoot}) coincides precisely with the ``root of baryonic Higgs branch'' on the moduli space. 

On the other side of the correspondence, we now have: 

{\bf Theory II:} Three dimensional $\cN=2$ supersymmetric gauge theory on $R^2\times S^1$ whose gauge group is $U(K)$,  and matter contents consist of $\Nc$ fundamental chiral multiplets with masses $M_l,~ l=1,\dots \Nc$;  $\Nc$ anti-fundamental chiral multiplets with twisted masses $\tM_l,~ l=1,\dots, \Nc$; and an adjoint chiral multiplets with twisted mass $\epsilon$.
This theory also has a FI parameter $r$ and a theta angle $\theta$ which can be combined to form a complex parameter $\hat{\tau}=ir+\frac{\theta}{2\pi}$, and for later purpose we also define $\hat{q}=e^{2\pi i \hat{\tau}}$. 

The F-term vacua of Theory II can be extracted by minimizing the one loop exact effective twisted superpotential calculated using the results in \cite{Nekrasov2009A}, and they are given by the Bethe Ansatz equation of inhomogeneous twisted XXZ spin chain:
\begin{equation}\label{BAE}
 \prod_{k=1}^{N_c}\frac{\sinh\left(\frac{R(\lambda_{li}-M_{k})}{2}\right)}{\sinh\left(\frac{R(\lambda_{li}-\tilde{M}_{k})}{2}\right)} 
 = -q \prod_{j=1}^{N_c} \prod_{k=1}^{n_k}\frac{\sinh\left(\frac{R(\lambda_{li}-\lambda_{kj}-\ep)}{2}\right)}{\sinh\left(\frac{R(\lambda_{li}-\lambda_{kj}+\ep)}{2}\right)}.
 \end{equation}
Here $q=e^{2\pi i\tau}=(-1)^{N_c+1} \hat{q}$ and $\{\lambda_{li}\}$ are the vevs of the adjoint scalar in the vector multiplet, and  we have introduced the double index notation $li$, where $\{n_l\}$ satisfy $\sum_{l=1}^{N_c} n_l = K$, this can be most easily understood from the D-brane picture that the theory is now in the Higgs phase and we are distributing $K$ different vevs among $N_c$ different mass parameters $\{M_l\}$ \cite{DHL}, $n_l$ is the number of the vevs associated with $M_l$. 

This three dimensional theory on $R^2\times S^1$ is precisely identified as the world volume theory of the BPS vortices living in  the Theory I, 
which can be regarded as the ultra-violet limit of the co-dimension two surface operators featuring prominently in the various recent studies on 2d/4d correspondence, beginning with \cite{AGGTV, AT2010}. In contrast with the non-dynamical surface operators, which are insertions in the gauge theories, the vortices exhibit rich world volume dynamics and are known to capture the BPS spectrum of the underlying supersymmetric gauge theory where they are embedded \cite{TongReview}. 
Crucially  such non-trivial dynamics also allow for the exact correspondence and its interpretation as a realization of refined geometric transition, that we shall review next.
\paragraph{}
To extend the exact correspondence in \cite{DHL, CDHL} which gives a one to one map between the chiral rings of certain four and two dimensional supersymmetric gauge theories, our starting point is the instanton partition function for Theory I computed by localization techniques in \cite{Nekrasov2002, Nekrasov2003} .
This amounts to topologically twisting the gauge theory on Omega-deformed background and summing over the fixed points of the action of cohomological charges.  The resultant expression is labeled by a set of Young diagrams.
There are many equivalent ways to express the instanton partition function, thanks to the series of useful identities proven in  \cite{Awata2008}.  We shall make use of many of them extensively throughout this note and refer interested readers to the proofs in their original paper. Here we begin by expressing it in terms of products of Pochhammer symbol \cite{Awata2008, Awata2010a}:
\begin{equation}
\Zinst(Q; q, t)=\sum_{\{\vec{Y}\}}\prod_{l,k=1}^{N_c}\left[\frac{\Lambda^{2|Y_l|}}{v^{|Y_l|}}\left(\frac{Q_{k}^{+}}{Q^{-}_k}\right)^{\frac{|Y_l|}{2}}
\frac{\cN_{\Yl 0}\left(v\frac{Q_l}{Q_k^{+}}; q, t\right)\cN_{ 0 \Yl }\left(v\frac{Q_k^{-}}{Q_l}; q, t\right)}{\cN_{\Yl \Yk}\left(\frac{Q_l}{Q_k}; q, t\right)}\right]\,,\label{Zinst5D}
\end{equation}
where $\vec{Y}=(Y_1,Y_2,\dots, Y_{N_c})$ is a set of Young diagrams whose columns satisfy $Y_{l1}\ge Y_{l2}\ge \dots $, for the time being the number of the columns in each Young diagram can be arbitrary. 
The various quantities in this expression are defined as:
\begin{eqnarray}\label{Notations}
&&(q, t)=(e^{R\ep_2}, e^{-R\ep_1}), \quad  (u, v)= (\sqrt{qt}, \sqrt{q/t}), \quad  (v Q_l,  Q_l^{+},  Q_l^{-})=(e^{Ra_l}, e^{R m_l}, e^{R \tm_l}), \nonumber
\\
\end{eqnarray}
\begin{eqnarray}
\cN_{\lambda\mu}(Q; q,t)&=&\prod_{i,j=1}^{\infty}\frac{\left(Q q^{\lambda_i-\mu_j} t^{j-i+1}; q\right)_{\infty}}{\left(Q q^{\lambda_i-\mu_j} t^{j-i}; q\right)_{\infty}}
\frac{\left(Q  t^{j-i}; q\right)_{\infty}}{\left(Q  t^{j-i+1}; q\right)_{\infty}},\nonumber\\
&=&\prod_{(i, j)\in \Yl}\left(1-Q q^{\lambda_i-j} t^{\mu_j^{\rm T}-i+1}\right) \prod_{(i,j)\in \mu}\left(1-Q q^{-\mu_{ i}+j-1} t^{-\lambda_{j}^{\rm T}+i}\right),\nonumber\\
&=& \prod_{(i, j)\in \Yk}\left(1-Q q^{\lambda_i-j} t^{\mu_j^{\rm T}-i+1}\right) \prod_{(i,j)\in \Yl}\left(1-Q q^{-\mu_{i}+j-1} t^{-\lambda_{j}^{\rm T}+i}\right). \label{DefcN}
\end{eqnarray}
with $(Z; q)_L=\prod^{L-1}_{r=0}(1-Z q^r)$ the Pochhammer symbol and $R$ the compactification radius.
To take the NS-limit $(\epsilon_1,\epsilon_2)\to (\epsilon, 0)$, we can apply the following identity :
\begin{equation}
\log (Z; q)_{\infty} = \frac{1}{\epsilon_2} \sum_{m=0}^{\infty} \frac{B_m (-\epsilon_2)^m}{m !} {\rm Li}_{2-m} (Z) + \log(1-Z)
\end{equation}
where $B_m$ are the Bernoulli numbers. The instanton partition function (\ref{Zinst5D}) at leading order of $\epsilon_2$ expansion can now be written as:
\begin{equation}\label{ZinstNS}
\Zinst(Q; (\epsilon_1, \epsilon_2) \to (\epsilon, 0))=\sum_{\{\vec{Y}\}}\exp\left[\frac{1}{\epsilon_2} \cH_{\rm inst.}\left(X_{li}, X_{li}^{(0)}\right) \right],
\end{equation} 
where $\cH_{\rm inst.}\left(X_{li}, X_{li}^{(0)}\right) = \cY(X_{li},X_{kj})-\cY\left(X_{li}^{(0)}, X_{kj}^{(0)}\right)$,
\begin{eqnarray}
\cY(X_{li}, X_{kj})&=&\sum_{li} X_{li} \log {\tilde{\Lambda}}^2 +\frac{1}{2}\sum_{li}\sum_{kj}\left[{\rm Li}_2\left(e^{R(X_{li}-X_{kj}+\epsilon)}\right)-{\rm Li}_2(e^{R(X_{li}-X_{kj}-\epsilon}))\right]\nonumber\\
&-& \sum_{li}\sum_{k}\left[{\rm Li}_2\left(e^{R(X_{li}-m_k)}\right)+{\rm Li}_2\left(e^{R(X_{li}-\tm_k-\ep)}\right)\right]
\end{eqnarray}
and $\cY\left(X_{li}^{(0)}, X_{kj}^{(0)}\right)=\cY\left(X_{li} \to X_{li}^{(0)}, X_{kj} \to X_{kj}^{(0)}\right)$ \footnote{Here we have also defined the modified dynamical scale:
$\tilde{\Lambda}^2 = \frac{\Lambda^{2N_c}}{v^{N_c}} \exp\left(\frac{R}{2}\sum_{k=1}^{N_c} (m_k-\tm_k)\right)$.}. 
The new variables $\{X_{li}\}$ and $\{X_{li}^{(0)}\}$ are defined as:
\begin{equation}
X_{li}=a_l + i \epsilon + Y_{li} \epsilon_2, \quad X_{li}^{(0)}=a_l+i \epsilon, \quad l=1,\dots N_c, ~i=1, \dots \infty,
\end{equation}
Notice that we are keeping the term $Y_{li}\ep_2$ which can be finite, as the length of the column $Y_{li}$ can still be infinite in the $\ep_2\to 0$ limit.
Also in this limit the discrete $\{X_{li}\}$ and $\{X_{li}^{(0)}\}$ become continuous distributions and using the standard matrix model techniques we can rewrite the Hamiltonian $\cH_{\rm inst.}$ into the following functional:
\begin{equation}\label{Ham1}
\cH_{\rm inst.}\left(X_{li}, X_{li}^{(0)}\right) = -\frac{1}{2} \int_{\cJ\times \cJ} dx dy \rho(x)G(x-y)\rho(y) +\int_{\cJ} dx \rho(x) \log \left[\tilde{q} R(x) \right]
\end{equation}
Here the various quantities are defined as:
\begin{equation}
G(x)=\frac{d}{d x} \log\frac{\sinh \left(\frac{x-\ep}{2}\right)}{\sinh \left(\frac{x+\ep}{2}\right)},\quad R(x)= \frac{A(x-\ep) D(x)}{P(x-\ep)P(x)}, 
\quad \tilde{q}=\tilde{\Lambda}^2,
\end{equation}
\begin{equation}\label{funcs}
A(x)=\prod^{N_c}_{k=1}\sinh \left(\frac{R(x-\tm_k)}{2}\right), \quad D(x)=\prod^{N_c}_{k=1}\sinh \left(\frac{R(x-m_k)}{2}\right),\quad 
P(x)=\prod^{N_c}_{k=1}\sinh \left(\frac{R(x-a_k)}{2}\right),
 \end{equation}
and $\rho(x)$ is the density function which is constant and only non-vanishing on the the cuts  $\cJ=\bigcup_{li} \left[ X_{li}^{(0)}, X_{li}\right]$ formed by the condensation of $\{X_{li}\}$. 
 
In the NS limit, the partition function $\Zinst$ is dominated by an infinite set of saddle points, as can be seen by the variation of density function $\rho(x) $ or equivalently the end points $\{X_{li}\}$. However it was shown in \cite{CDHL} that we can truncate such infinite set by imposing the quantization condition on the cycles corresponding to the undeformed Higgs branch root condition (\ref{HiggsRoot}).  
The quantization condition is equivalent to truncate the Young diagram $Y_{li}$ to only $n_l$ columns, i.e. 
 \begin{equation}\label{TruncX}
 {\rm For}~ i > n_l \quad  X_{li}=X_{li}^{(0)}, \quad \longleftrightarrow  \quad Y_{li}=0\,.
 \end{equation}
This truncation of Young diagrams can also be easily deduced from the second and third lines in the definition of $\cN_{Y_l Y_k}(Q; q,t)$ (\ref{DefcN}) even before taking $\ep_2\to 0$ limit, to have non-vanishing partition function $\Zinst$ when we impose $vQ_{l}=Q_{l}^{+} t^{n_l}$, it is necessary for $Y_{li}$ to have at most $n_l$ columns.
The resultant set of $\{X_{li}\}$ satisfies the following equation:
\begin{equation}\label{BAE1}
\frac{D(X_{li}+\ep)}{A(X_{li}-\ep)}=-\tilde{q}\frac{S(X_{li}-\ep)}{S(X_{li}+\ep)}, \quad S(X)=\prod_{k=1}^{N_c}\prod_{j=1}^{n_l}\sinh\left(\frac{R(x-X_{kj})}{2}\right).
\end{equation}
We recognize that up to re-definition of parameters, this is precisely the Bethe Ansatz Equation for inhomogeneous twisted XXZ spin chain, in other words, the discrete vacua of Theory I in NS limit are in one to one correspondence with the eigenvectors of the twisted XXZ spin chain Hamiltonian.
We will provide an alternative derivation for this equation from the difference equation obeyed by the truncated instanton partition function momentarily. 
We can connect this result with Theory II by making the following identification between the two sets of  parameters:
\begin{equation}\label{matching1}
X_{li}=\lambda_{li}+\frac{1}{2}\ep, \quad m_l=M_l+\frac{3}{2} \ep, \quad \tm_l=\tilde{M}_l-\frac{1}{2}\ep. 
\end{equation}
As the equation (\ref{BAE1}) reduces to the one in (\ref{BAE}) for Theory II, this establishes a precise one to one map between the vacua of Theory I and Theory II.
One can also show that given the quantization condition (\ref{TruncX}) and the matching of parameters (\ref{matching1}), the Yang-Yang generating functional of the twisted XXZ spin (\ref{BAE}),  which is the twisted superpotential of Theory II, coincides with the functional (\ref{Ham1}) up to a perturbative piece. This is taken care of by the perturbative part of Nekrasov partition function for Theory I (see \cite{CDHL} for more details).

\paragraph{}
The saddle point analysis we have reviewed so far allows us to establish an exact correspondence between the vacua of the compactified 5d and 3d gauge theories. 
Here we also study the difference equation obeyed by the truncated instanton partition function, and this provides an alternative derivation of the Bethe Ansatz Equation (\ref{BAE1}). The same difference equation also appears in the refined open topological string computation, from which we can derive the mirror curve for the associated toric geometries. We give an explicit illustration for this in Appendix B.
Using the identity for Pochhammer symbols:
\begin{equation}
(a ; q)_L= \frac{(a ; q)_\infty}{(a q^L ; q)_\infty}, \label{PochId}
\end{equation}
we can further rewrite the instanton partition with truncated Young diagrams into
\begin{equation}
\Zinst(Q; q, t)=\sum_{\{\vec{Y}\}_{\vec{n}}}\prod_{l,k=1}^{N_c}\left[ {U_k}^{|Y_l|}
\prod_{i=1}^{n_l}
\frac{\left(e^{-X_{li}+\tm_k} ; q\right)_{Y_{li}}}{\left(e^{-X_{li}+m_k}; q\right)_{Y_{li}}}\prod_{i=1}^{n_l}\prod_{j=1}^{n_k}\frac{\left(e^{X_{li}-X_{kj}}; q\right)_{Y_{kj}-Y_{li}}}{\left(e^{X_{li}-X_{kj}}t; q\right)_{Y_{kj}-Y_{li}}}\right]\label{trZpoch}.
\end{equation}
Here we have also defined the variable
\begin{equation}
U_k = \frac{\Lambda^2}{v^{1/2}} \left( \frac{Q_k^{+}}{ Q_k^{-}}\right)^{1/2},
\end{equation}
and $\tilde{\Lambda}^{2}=\prod_{k=1}^{N_c} U_k$.
The summation over $\{\vec{Y}\}_{\vec{n}}$ includes all possible vectors $\vec{n}=(n_1, n_2, \dots, n_l,\dots, n_{N_c})$, and for each Young-diagrams of $n_l$ columns, we sum over their all possible lengths i. e. $Y_{li}=1, 2, \dots, \infty$. 
Now if we consider for given $\vec{n}$, the difference between the two terms in (\ref{trZpoch}) with $Y_{l'i'}$ and $Y_{l'i'}+1$ while all other $Y_{li}, ~ \{li\} \neq \{l'i'\}$ remain fixed and equal, we can derive that, after various cancelations, the two such terms differ by an overall factor:
\begin{equation}
\prod_{k=1}^{N_c}\left\{ U_k \frac{(1-e^{-X_{l'i'}+\tm_k} q^{-1})}{(1-e^{-X_{l'i'}+m_k}q^{-1})}\prod_{j=1}^{n_k}\frac{(1-e^{X_{l'i'}-X_{kj}}t)(1-e^{-X_{l'i'}+X_{kj}}q^{-1})}{(1-e^{-X_{l'i'}+X_{kj}}t)(1-e^{X_{l'i'}-X_{kj}})}\right\}.\label{extraFactor}
\end{equation}
We can also define the shift operators for $U_b$:
\begin{equation}
T_{t, U_k}= t^{U_k\partial_{U_k}}=e^{-\epsilon_1 U_k\partial_{U_k}}
\quad
T_{q^{-1}, U_k}=q^{-U_k\partial_{U_k}}=e^{-\epsilon_2 U_k\partial_{U_k}},
\end{equation}
such that $T_{t, U_k} U_k =t U_k$ and $T_{q^{-1}, U_k}U_k = q^{-1} U_k$.
The extra factor  (\ref{extraFactor}) in fact equals to unity, since we are summing over Young diagrams with all column lengths. We can then readily write down the difference equation for the truncated instanton  partition function (\ref{trZpoch}) which obey:
\begin{eqnarray}\label{DiffEqn1}
&&\prod_{k=1}^{N_c}{\Big\{ }(1-e^{-X_{li}+m_k} T_{q^{-1}, U_k})\prod_{j=1}^{n_k}(1-e^{X_{li}-X_{kj}})(1-e^{-X_{li}+X_{kj}}T_{t, U_k})\nonumber\\
&& - U_k (1-e^{-X_{li}+\tm_k} T_{q^{-1}, U_k})\prod_{j=1}^{n_k}(1-e^{X_{li}-X_{kj}}T_{t, U_k})(1-e^{-X_{li}+X_{kj}}T_{q^{-1}, U_k}){\Big\}}\Zinst(Q;q,t)=0.\nonumber\\
\end{eqnarray}
The Bethe Ansatz Equation for twisted XXZ spin chain can be readily recovered in the NS limit $(\ep_1, \ep_2) \to(\ep, 0)$ from (\ref{extraFactor}) , equivalently we can also recover the Yang-Baxter equation by considering the following transfer function
\begin{eqnarray}
T(X)=\frac{1}{S(X)}\left\{D(X)S(X+\epsilon)-\tilde{\Lambda}^2 A(X) S(X-\epsilon)\right\}
\label{DefTx},
\end{eqnarray}
where various functions are defined in (\ref{funcs}).
Notice that the simple poles at $X=X_{li}$ from $1/S(X)$ in (\ref{DefTx}) are precisely canceled by $\{\dots \}=0$ at $X=X_{li}$ in the numerator, as implied by setting (\ref{extraFactor}) equals to one,  and $T(X=X_{li})$ remains finite. This allows us to deduce the following quantum Yang-Baxter equation:
\begin{eqnarray}
&&\frac{1}{S(X_{li})}\left\{D(X_{li})S(X_{li}+\epsilon U_k\partial_{U_k})-\tilde{\Lambda}^2 A(X_{li}) S(X_{li}-\epsilon U_k\partial_{U_k})\right\} \Zinst(Q; q,t)\nonumber\\
&&= T(X_{li}) \Zinst(Q; q,t) = \prod_{k=1}^{N_c}\sinh\left(\frac{R(X_{li}-a_k)}{2}\right)\Zinst(Q; q,t)\,.\label{XXZYBE}
\end{eqnarray}
We can interpret this as quantized version of associated five dimensional Seiberg-Witten curve, and the expansion coefficients for the monomials of $X_{li}$ should be interpreted as the commuting quantum conserved charge acting on the instanton partition function.
We will later see that the truncated Nekrasov instanton partition function considered here can be re-interpreted as refined open topological string amplitudes, i.e. containing explicit toric brane insertions, through a refined version of geometric transition. Open topological string amplitudes are known to satisfy certain difference equation even in the unrefined limit and from this we can also read off the corresponding dual quantum mirror curve/geometry whose coordinates are promoted to differential operators \cite{Aganagic2003}. The difference equation we derive here (\ref{DiffEqn1}) serves precisely as such an example for the corresponding toric geometry with multiple toric brane insertions. 
The difference equation for the simplest case with single toric brane insertion and single non-vanishing equivariant parameter has been derived earlier in \cite{Taki2010}, we generalize this in Appendix B to the case when both equivariant parameters are non-vanishing. 
 
\begin{figure}
\centering
\includegraphics[width=100mm]{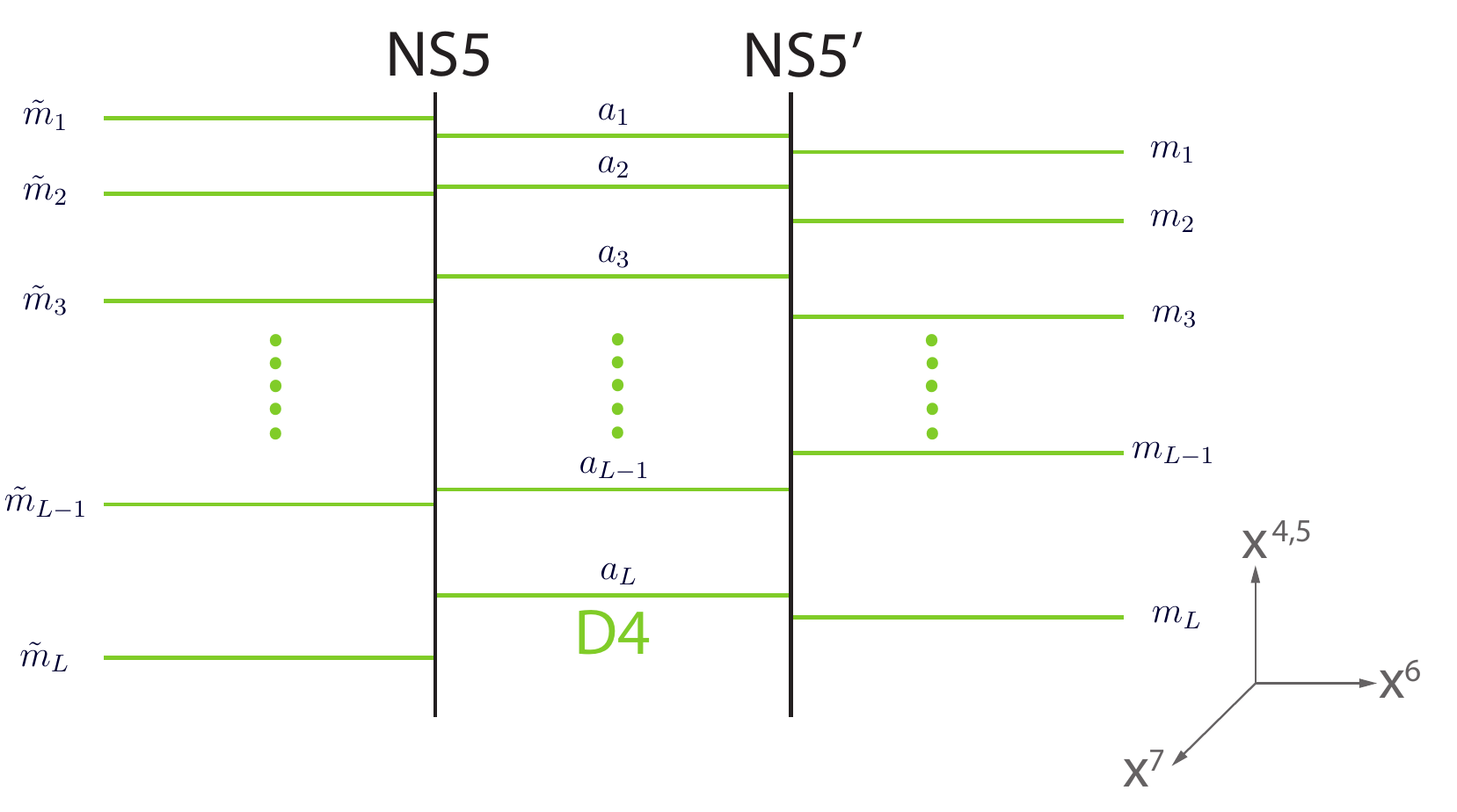}
\caption{The IIA-brane construction for Four Dimensional Limit of Theory I }
\label{THEORYI}
\end{figure}
\begin{figure}
\centering
\includegraphics[width=100mm]{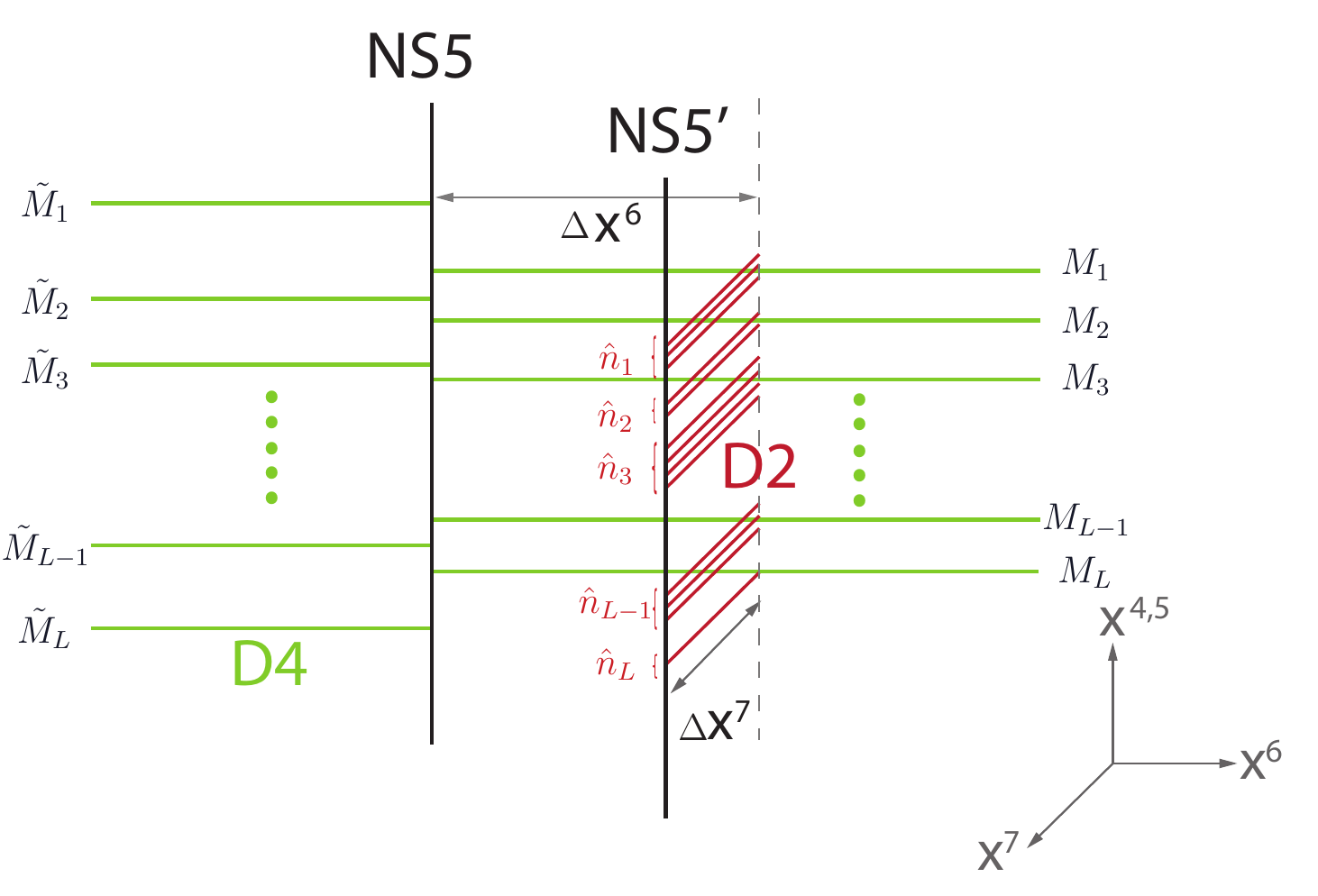}
\caption{The IIA-brane brane construction for Two Dimensional Limit of Theory II }

\label{THEORYII}
\end{figure}

Let us discuss the connections between the two field theories considered here in more details.
Generically when the co-dimension two vortices/surface operators are introduced into the underlying gauge theory (such as our Theory I), 
we can classify the non-perturbative effects in the resultant partition function into two different types. First we have the Yang-Mills instantons given by the self-dual field strength in the four/five dimensional gauge theory, but now have singularities along the world volume of the vortices\footnote{More precisely what we mean by five dimensional instantons come from usual four dimensional instantons now smearing over compactified $S^1$, i.e. an instantonic string.}.
In addition we can now have the ``sigma model lumps'' \cite{Atiyah1984} or ``vortices within vortices'' using the terminology in \cite{Hanany2004},
they are constructed from the field strength which is restricted to the vortex world volume, characterized by the topologically non-trivial map from the vortex world sheet onto the two cycles in the vortex moduli space. The quantization condition (\ref{HiggsRoot}) with $n_l \neq 0$ precisely corresponds to the limit where four/five dimensional coupling constant vanishes and simultaneously introducing vortices/sigma model lumps on the vortices\footnote{In contrast if we only have the classical baryonic Higgs condition $a_l=m_l$, this would only corresponds to decoupling the four/five dimension dynamics, and we would still need to put in vortices by hand.}, therefore the bulk instantons decouple from the partition function, and the non-perturbative contributions come solely from sigma model lumps in the vortex world volume. This decoupling can be most easily seen from the D-brane pictures provided in Figures \ref{THEORYI} and \ref{THEORYII} \footnote{We are ignoring the compactified $S^1 :$ $x^5 \sim x^5+2\pi R$ for the time being, which can be incorporated through smearing configuration along the $x^5$ direction.}, at the root of baryonic Higgs branch, the NS5 branes can move off in the transverse 789 directions and the gauge D4 branes become semi-infinite, and effective four dimensional gauge coupling vanishes.  
The BPS spectra of the vortex sigma model are quantum corrected by these sigma model lumps, and can be captured by the one loop exact twisted superpotential, as can be most easily seen from the weak coupling expansions  \cite{Dorey1998, Dorey1999, Shifman2004}.  
In this note, we apply the alternative approach of geometric engineering \cite{HIV} to realize these closely related gauge theories of different dimensionalities in refined topological string theory. The corresponding amplitudes yields their respective partition functions, and refined geometric transition allows us to connect them.


\section{Truncated Partition Function and Surface Operators}
\begin{figure}
\centering
\includegraphics[width=65mm]{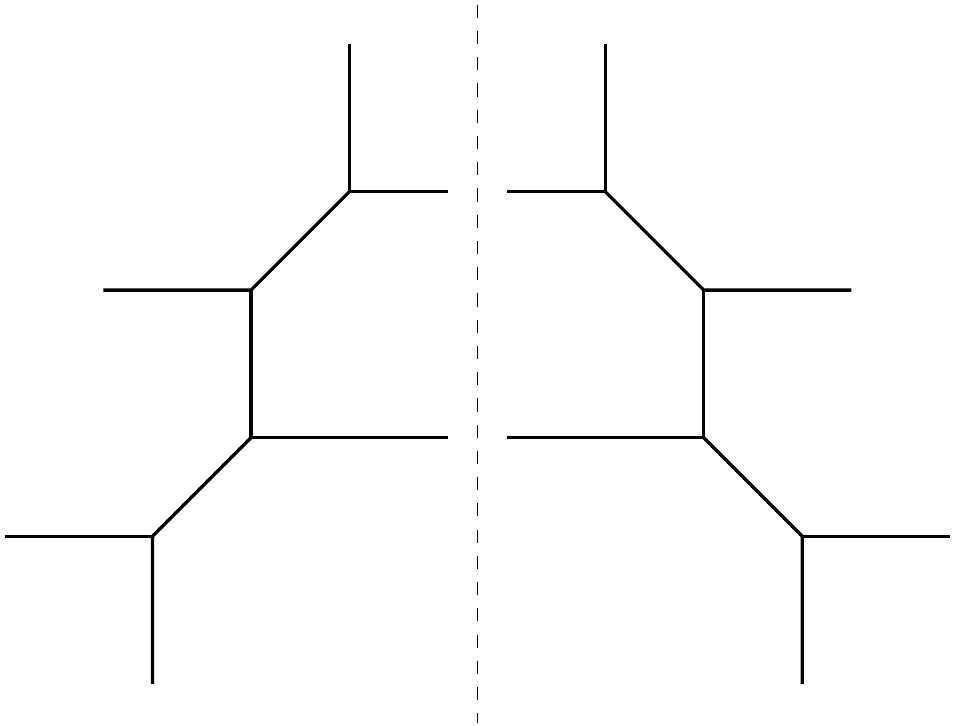}
\caption{The full glued geometry.}
\label{fullgeometry}
\end{figure}

\paragraph{}
In this section, we shall first restore both equivariant parameters $\epsilon_{1,2}$ and rewrite the truncated instanton partition function for Theory I (\ref{trZpoch}) into a form consisting few distinct parts, such that it is rendered suitable for subsequent discussion about the connections between our exact field theoretic correspondence and refined topological string computations. We argue that the general truncated instanton partition function can be readily reproduced by normalized refined topological string amplitudes with toric brane insertions, which in turns can be identified with gauge theory instanton partition function in the presence of surface operators.
In particular, the crucial quantization condition (\ref{HiggsRoot}) is realized in refined topological A-model as the condition where the corresponding K\"ahler moduli in the corresponding toric geometry engineering the gauge theory become degenerate, i.e. shrinking to zero size. Such geometric singularities can be resolved by massless toric branes wrapping on the dual Lagrangian three cycles, this is precisely a realization of geometric transition or so-called ``bubbling'' \cite{Gomis2007}.
\paragraph{}
It has been shown in \cite{Taki2007, Awata2008} that the instanton partition function for Theory I  (\ref{Zinst5D}) can be reproduced from refined closed topological string partition function, the corresponding toric Calabi-Yau can be obtained by the successive blow-up of $A_{N_c-1}$ ALE space fibered over $\bP^1$, see Figure \ref{fullgeometry} for the simplest resultant toric diagram with $N_c=2$. We can identify gauge theory parameters in such geometry, the K\"ahler modulus/area of the base $\bP^1$ gives the coupling/dynamical scales of the gauge theory,  the K\"ahler moduli of the fibers or vertical edges give the difference between the Coulomb vev, and the K\"ahler moduli for the blown-up $\bP^1$s or the tilted edges give the masses of the matter fields.  As shown in \cite{HIV}, M-theory compactified on such toric Calabi-Yau times $S^1$ gives an equivalent and complementary realization of Theory I to the D-brane set up discussed earlier, in particular the instantons are now realized as M2 branes wrapping over M-Theory $S^1$ and base $\bP^1$ in the toric Calabi-Yau, whose mass is now given by the sum of gauge coupling and KK-momentum. To compute the refined topological string amplitude on such toric Calabi-Yau, so-called ``refined topological vertex''  was introduced and employed \cite{IKV}. 
Explicitly we cut the toric diagram horizontally (``preferred direction'') into left and right strips (so-called ``strip geometry'') and assign a Young diagram to each of the three toric legs meeting at every refined vertex \cite{Iqbal2004},  in particular the total number of boxes in the Young diagrams for the horizontal legs give the instanton number in the resultant gauge theory. These Young diagrams specify the refined topological vertices for a strip geometry and the partition function can be obtained by multiplying together different vertices and summing over the intermediate Young diagrams, and glue together the left and right strips.

\begin{figure}
\centering
\includegraphics[width=70mm]{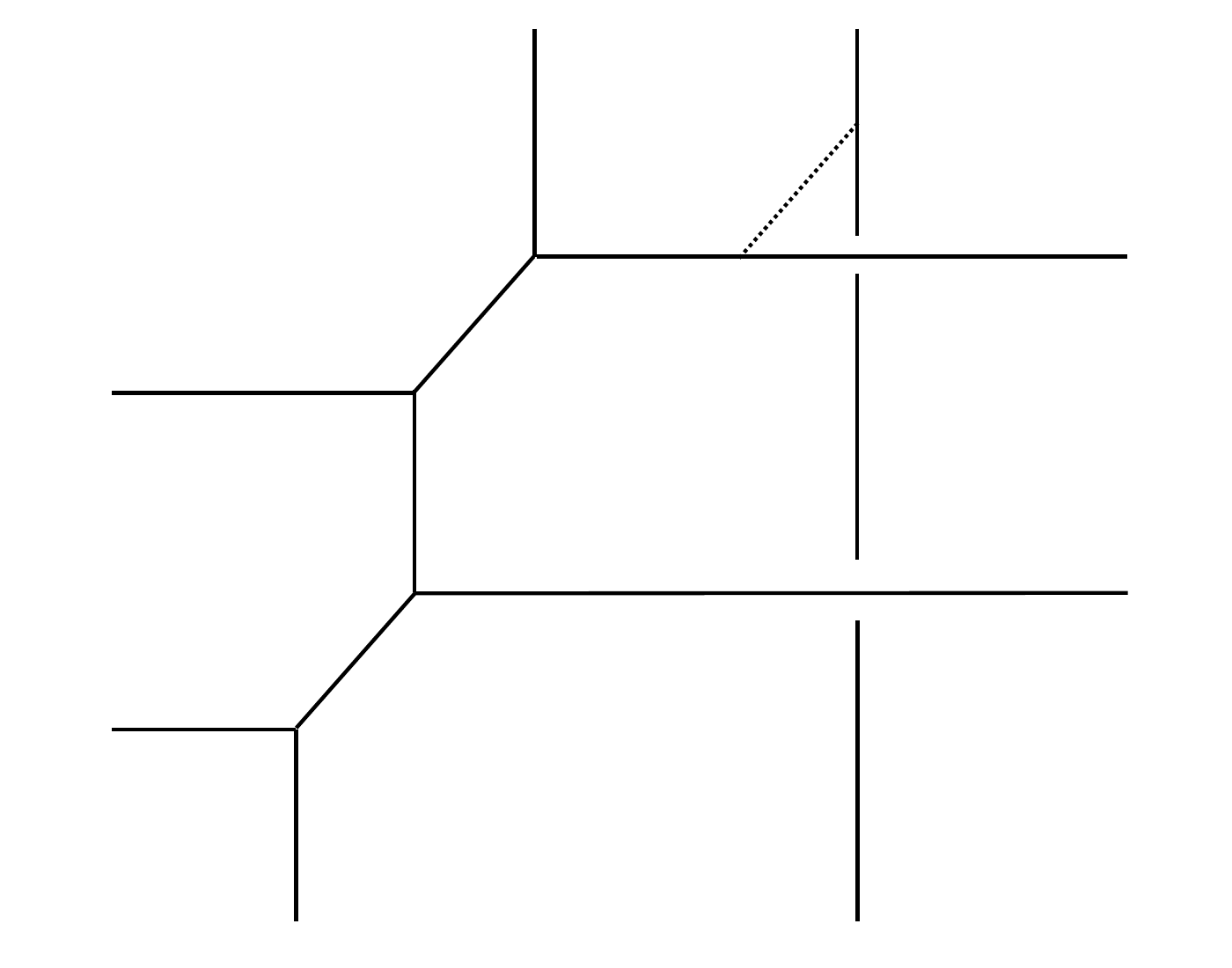}
\caption{Single Surface Operator Insertion}
\label{1surface}
\end{figure}

For the time being however, we would like to first discuss the connections between our exact field theory correspondence and the generalization of geometric transition in refined topological strings. Let us consider the most general truncated instanton partition function (\ref{Zinst5D}) with Young diagrams $Y_l$, $l=1,\dots, N_c$ containing finite $n_l \in {\mathbb Z}$ columns.  Here we can use another representation of the function $\cN_{Y_l Y_k}(Q; q, t)$ and its symmetry property as proven in \cite{Awata2008},
 \begin{equation}
 {\mathcal N}_{Y_l Y_k}(Q; q, t)=\prod_{(i,j)=1}^{\infty}\frac{1-Q q^{Y_{li}-j}t^{Y_{kj}^{\rm T}+1-i}}{1-Q q^{-j}t^{1-i}}
 =\prod_{(i,j)=1}^{\infty}\frac{1-Q q^{i-Y_{kj}-1}t^{j-Y_{li}^{\rm T}}}{1-Q q^{i-1}t^{j}}\,,
\end{equation}
to rewrite (\ref{Zinst5D}) into a suitable form for comparison with the open refined topological string computation. 
\begin{figure}
\centering
\includegraphics[width=40mm]{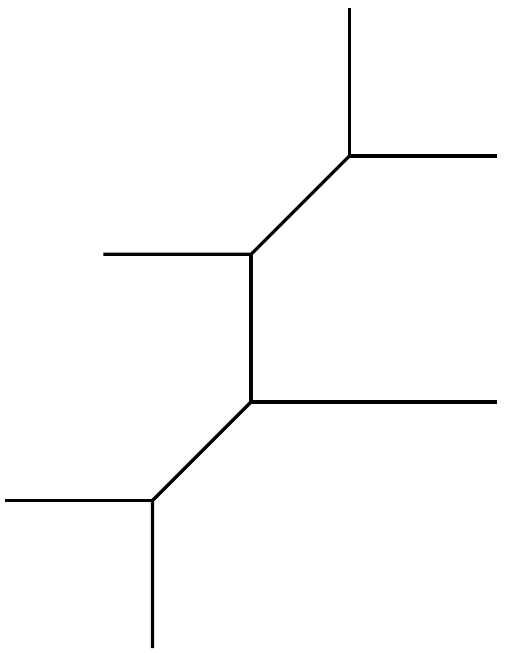}
\caption{The left strip geometry with fixed partitions}
\label{halfgeometry}
\end{figure}
After some straightforward albeit tedious rewriting, we can decompose the truncated instanton partition function as:
\begin{eqnarray}\label{ZinstTop}
&&\Zinst(Q; q,t)
=\sum_{\{\vec{Y}\}_{\vec{n}}}\prod_{l=1}^{N_c}  q^{\frac{1}{2}||Y_l ||^2} \tilde{\Lambda}^{2|Y_l|} P_{Y_l^{\rm T}}(q^{\rho}, t^{n_l} q^{\rho}; t, q)\times {\mathcal Z}_{\rm frame}\times  {\mathcal Z}_{\rm Ver.}
\times {\mathcal Z}_{\rm OV}.
\end{eqnarray}
Here $||Y_l ||^2=\sum_{i=1}^{n_l}Y_{li}^2$,   $P_{Y_l^{\rm T}}(q^{\rho}, t^{n_l}q^{\rho}; t, q)$ is the principal specialization of MacDonald polynomial given in the Appendix B of \cite{Awata2008}, $x q^{\rho} \equiv x q^{\frac{1}{2}-i}, i=1, 2\dots \infty$.
In the decomposition (\ref{ZinstTop}), we have also defined:
\begin{eqnarray}\label{ZVer}
{\mathcal Z}_{\rm Ver.}&=&\prod_{l > k}^{N_c} \prod_{(i,j)=1}^{\infty}\frac{1-\frac{Q_k}{Q_l} q^{j-1}t^{i}} {1-\frac{Q_k}{Q_l} q^{j-Y_{li}-1}t^{i-Y_{kj}^{\rm T}}}
\prod_{k,l =1}^{N_c}\prod_{(i,j)=1}^{\infty}\frac{1-\frac{Q_k^-}{Q_l v} q^{j-1-Y_{li}}t^{i}} {1-\frac{Q_k^-}{Q_l v} q^{j-1}t^{i}},\nonumber\\
&=&\prod_{l > k}^{N_c} \prod_{(i,j)=1}^{\infty}\frac{1-\frac{Q_k}{Q_l} q^{-i}t^{1-j}} {1-\frac{Q_k}{Q_l} q^{Y_{kj}-i}t^{Y_{li}^{\rm T}+1-j}}
\prod_{k,l =1}^{N_c}\prod_{(i,j)=1}^{\infty}\frac{1-\frac{Q_k^-}{Q_l v} q^{j-1-Y_{li}}t^{i}} {1-\frac{Q_k^-}{Q_l v} q^{j-1}t^{i}},
\end{eqnarray}
\begin{eqnarray}
{\mathcal Z}_{\rm OV}&=&\prod_{l > k}^{N_c} \prod_{(i,j)=1}^{\infty}\frac{1-\frac{Q_l}{Q_k} q^{-j}t^{1-i}}{1-\frac{Q_l}{Q_k} q^{Y_{li}-j}t^{Y_{kj}^{\rm T}+1-i}} 
 \prod_{l\neq k}^{N_c}\prod_{i,j=1}^{\infty}\frac{1-v\frac{Q_l}{Q_k^+} q^{Y_{li}-j}t^{1-i}}{1-v\frac{Q_l}{Q_k^+} q^{-j}t^{1-i}},\nonumber \\
 &=&\prod_{l>k}^{N_c}\prod_{(i,j)}\frac{(\frac{Q_l}{Q_k}q^{Y_{li}-Y_{kj}}t^{j-i};q)_{\infty}}{(\frac{Q_l}{Q_k}q^{Y_{li}-Y_{kj}}t^{j-i+1};q)_{\infty}}
 \frac{(\frac{Q_l}{Q_k}t^{j-i+1};q)_{\infty}}{(\frac{Q_l}{Q_k}t^{j-i};q)_{\infty}} 
\prod_{l\neq k}^{N_c}\prod_{(i,j)} \frac{1-v\frac{Q_l}{Q_k^+}q^{Y_{li}-j}t^{1-i}}{1-v\frac{Q_l}{Q_k^+}q^{-j}t^{1-i}},
  \label{ZOV}
 \end{eqnarray}
and 
\begin{equation}\label{Zframe}
{\mathcal Z}_{\rm frame} = \prod^{N_c}_{l=1}\prod_{i=1}^{n_l}\prod_{j=1}^{Y_{li}}\frac{1}{1-q^{j-Y_{li}-1}t^{i-Y_{lj}^{\rm T}}}.
\end{equation}
Now we would like to discuss how these various components in the truncated instanton partition or closed refined topological string partition function (\ref{ZinstTop}) can also arise from refined open topological string partition function,  in other words the refined version of geometric transition.
In the M-theoretic construction of refined A-model open topological string, the surface operators/vortices on $R^2\times S^1$ are realized as a probe M5 branes wrapping on certain Lagrangian three cycle in the corresponding toric Calabi-Yau (as indicated by dotted line in Figure \ref{1surface}.), and also stretch in the non-compact $R^2$ and M theory $S^1$. There are one form gauge fields inside these Lagrangian three cycles, arising from the M2 branes wrapping on M-theory $S^1$ and a holomorphic two cycle with boundary on these wrapped M5 branes. The K\"ahler modulus of the two cycle M2s wrap on gives them their masses and it is identified with the dynamical scale $\sim e^{2\pi i \hat{\tau}}$ of vortex effective theory on $S^1\times R^2$, they therefore precisely appear as the ``sigma model lumps'' running along the $S^1$ \cite{Dimofte2010}.

The MacDonald polynomial in the truncated instanton partition (\ref{ZinstTop}) precisely corresponds to the non-trivial holonomy or Wilson loop of the gauge field in the probe toric branes integrating along the boundary of the M2 brane, whose value is labeled by Young diagram $Y_{li}$. 
It is also important to note that only when the quantization condition $a_l=m_l-n_l\ep$ is imposed in our original five dimensional Nekrasov partition function (\ref{Zinst5D}), we can combine the contributions from the hyper-multiplets and vector multiplet into a single MacDonald polynomial function, and deduce its interpretation as the holonomy of probe brane gauge field.  This condition is precisely the degeneration of the K\"ahler moduli in the corresponding toric Calabi-Yau for refine closed topological string theory, where toric branes emerge. This is a first indication that we can reproduce (\ref{ZinstTop}) from a refined open topological string partition function, where different toric brane insertions are represented by their holonomies.

Turning on non-trivial holonomies/Wilson lines for the gauge fields on probe toric M5 branes also implies that they are separated into different stacks labeled by $l=1,\dots, N_c$, where different vevs $\{a_l\}$ give the positions for different stacks of toric branes now wrapping on different special Lagrangian three cycles.
This matches with the interpretation that these toric branes are realized as surface operators as labeled by $a_l$, and charged under different Abelian factors in the original five dimensional gauge theory. 

\begin{figure}
\centering
\includegraphics[width=70mm]{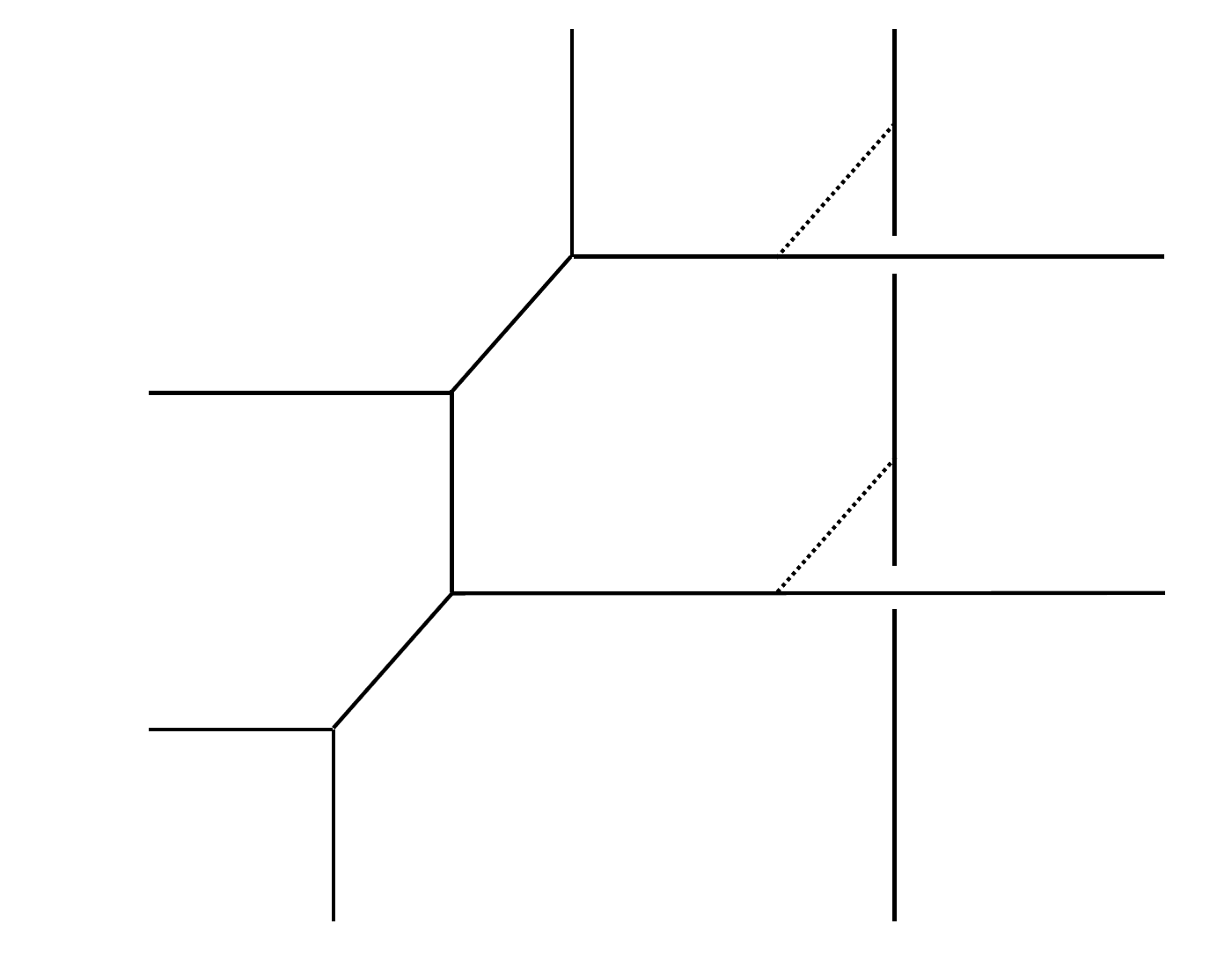}
\caption{Two Surface Operator Insertions}
\label{2surface}
\end{figure}

To apply the rules of refined topological vertex to configuration given in Figures \ref{1surface} and \ref{2surface}, we can again multiply together different refined topological vertices with given set of Young diagrams and sum over the intermediate diagrams, this precisely generates the factor $\cZ_{\rm Ver.}$ in (\ref{ZinstTop}) which corresponds to a strip geometry.  Different stacks of probe toric branes insertions are simply represented by their corresponding gauge holonomies,  i.e. MacDonald polynomials $\{ P_{Y_l^{\rm T}}(q^{\rho}, t^{n_l} q^{\rho}; t, q) \}$.   
While $\cZ_{\rm frame}$ is the normalized framing factor independent of the gauge theory parameters, this was introduced in \cite{Taki2007} to ensure the matching between the gauge theory and topological string computations. For a single stack of toric brane insertions, i.e. only one non-vanishing $n_l$, this would have been the entire story. Indeed explicit open refined topological string partition functions have been calculated in \cite{Dimofte2010, Taki2010}, and shown to be equal to closed refined topological string partition functions with degenerated moduli, establishing the simplest case of refined geometric transition.   
However we can have multiple stacks of toric branes and there are additional interactions between them, these are responsible for the generation of the remaining ${\mathcal Z}_{\rm OV} $ (\ref{ZOV}) in (\ref{ZinstTop}), which is the refined version of Ooguri-Vafa factor \cite{OVfactor},  from the corresponding refined open topological string partition function. 

Let us now discuss the origin of $\cZ_{\rm OV}$ in some details. As we now have multiple stacks of M5 branes wrapping distinct Lagrangian three cycles, we can also have M2 branes wrapping on a cylinder  $S^1 \times {\rm Interval}$ stretching between two of them, each one dimensional boundary of the annulus is charged under the gauge field in each stack of toric branes.  In general, the stretched M2 branes can again give massive modes, however ground state/minimal energy configuration occurs when the two Lagrangian three cycles intersects along a $S^1$ on the cylinder.  The refined Ooguri-Vafa factor emerges precisely from the one loop determinant generated by integrating out these light M2 branes modes charged under the gauge fields in the both Lagrangian three cycles. 
The resultant factor depends on the gauge holonomy in each stack of toric branes the M2 branes can end on, and again labeled by the corresponding Young diagrams. In the unrefined limit the explicit form has been calculated earlier in \cite{OVfactor, AMV2002}, while the refined generalization has also recently been derived in  \cite{ AS2011, AS2012}
\footnote{In addition we can regard the contribution from the hypermultiplets as gauge field in toric branes whose gauge coupling has been taken to zero, hence trivial holonomy.}. The key feature of these results is that the Ooguri-Vafa factor can be expressed in terms of the power sum for the traces of holonomy matrices. 

In fact we can see $\cZ_{\rm OV}$ can be written in such form by using the identity $(1-x)=\exp[\log(1-x)]=\exp[-\sum_{n=1}^{\infty} \frac{x^n}{n}], ~ |x|<1$ to obtain
\begin{eqnarray}
&&\exp\left[\sum_{l>k}^{N_c}\sum_{n=1}^{\infty}\frac{v^{-n}}{n}\left\{\sum_{i=1}^{n_l}\left(Q_l q^{Y_{li}}t^{1/2-i}\right)^n
 \sum_{j=1}^{n_k}\left(Q_k t^{-Y_{kj}^{\rm T}}q^{j-1/2}\right)^{-n} - \sum_{i=1}^{n_l}\left(Q_l t^{1/2-i}\right)^n
\sum_{j=1}^{n_k}\left(Q_k q^{j-1/2}\right)^{-n}\right\}\right]\nonumber\\
\times && \exp\left[-\sum_{l\neq k}^{N_c}\sum_{n=1}^{\infty}\frac{1}{n}\left\{\sum_{i=1}^{n_l}\left(Q_l q^{Y_{li}}t^{1/2-i}\right)^n
 \sum_{j=1}^{n_k}\left(Q_k^{+} q^{j-1/2}\right)^{-n}-\sum_{i=1}^{n_l}\left(Q_l t^{1/2-i}\right)^n
\sum_{j=1}^{n_k}\left(Q_k^{+} q^{j-1/2}\right)^{-n}\right\}  \right]\nonumber\\
\label{ZOV2}
\end{eqnarray}
Here the first line can be regarded as the one-loop determinant from integrating out M2 brane modes between two different stacks of toric branes, and $\sum_{i=1}^{n_l}\left(Q_l q^{Y_{li}}t^{1/2-i}\right)^n = {\rm Tr} (U^n_l)$ and $\sum_{j=1}^{n_k}\left(Q_k t^{-Y_{kj}^{\rm T}}q^{j-1/2}\right)^{-n} = {\rm Tr}(U_k)^{-n}$  are precisely the traces of the diagonal gauge holonomy matrix for each stack of toric branes as labeled by Young diagrams $Y_l$ and $Y_{k}$.
Similarly the second line of (\ref{ZOV2}) comes from integrating out each stack of toric branes and the background ``flavor branes'' with trivial gauge honolomy \footnote{The relative overall minus sign between the first and second lines of (\ref{ZOV2}) comes from the opposite orientations between the toric branes wrapping on the internal toric legs and the background flavor toric branes.}.
The double summation over empty Young diagrams in the second terms of both lines in (\ref{ZOV2}) are included here to obtain the relative normalization between the refined topological string amplitude and the gauge theory instanton partition function. 
Moreover using the matrix identity:
\begin{equation}\label{MatrixId}
\det\left[1-v^{-1} U\otimes V^{-1} \right]=\exp \left[{\rm Tr} \log(1-v^{-1}U\otimes V^{-1})\right]
=\exp\left[-\sum_{n=1}^{\infty}\frac{v^{-n}}{n} {\rm Tr} U^n {\rm Tr}V^{-n} \right],
\end{equation}
we can also rewrite the exponential summation into another familiar determinant form for the Ooguri-Vafa factor \cite{OVfactor, AMV2002, AS2012}.
Finally using the summation identity for the MacDonald polynomials:
\begin{equation}\label{MacSum}
\sum_{\lambda}P_{\lambda}(\{x_i\}; q, t) P_{\lambda^{\rm T}}(\{-y_j\}; t,q)=\prod_{i,j}(1-x_i y_j)=\exp\left[-\sum_{n=1}^{\infty} \frac{1}{n} p_n(x)p_n(y)\right],
\end{equation} 
where $p_n(x)=\sum_{i} x_i^n$, we can rewrite ${\mathcal Z}_{\rm OV}$ (\ref{ZOV2}) precisely into the form for refined Ooguri-Vafa factor derived recently in \cite{AS2012} (See equation (2.6) of \cite{AS2012} \footnote{Modulo a factor of dynamical scale which we have factored it out. }).  We conclude from the discussion above that $\cZ_{\rm OV}$ in (\ref{ZinstTop}) can be generated from the open refined topological string partition function by inserting the appropriate Ooguri-Vafa factor. Conversely, via refined geometric transition, we can predict the Ooguri-Vafa factor from the degeneration of appropriate closed refined topological string partition function. 

In the original D-brane picture, to generate precisely such refined Ooguri-Vafa factor,  we need to have all the D2 branes corresponding to surface operators ending on the same regulating NS5 brane, otherwise we would instead generate quiver gauge group in the vortex world volume theory; via geometric engineering this NS5 brane is precisely mapped to a single additional ``toric degeneration locus'' such that all the inserted toric branes can end on \cite{Dimofte2010} as indicated by the vertical line in Figures \ref{2surface}, the final result however does not depend on the position of the degenerated locus and can be taken to infinity. 
\paragraph{}
While we have not shown it explicitly here, let us comment about the relation between open topological string amplitudes and twisted superpotential of Theory II considered earlier. In \cite{Shadchin2006} (see also \cite{Yoshida}),  localization techniques were applied to compute the partition for a class of two dimensional $\cN=(2,2)$ supersymmetric gauge theories, the resultant partition consists of an one loop perturbative part and non-perturbative contributions which can be interpreted as summing over the world sheet instantons/vortices. The final expression can be expressed as the exponential of integral over the equivariant deformation of the twisted superpotential obtained in \cite{Witten1993}.  More precisely as explained in \cite{DHL} that we can identify the equivariant parameter $\epsilon_1$ transverse to the vortices as the adjoint hypermultiplet mass in vortex world volume theory, while $\epsilon_2$ which is set to zero in NS limit,  can be restored as the equivariant parameter in the vortex world volume theory when performing two dimensional localization computations (See \cite{Benini2012, Doroud2012, Gomis2012} for more details).
Moreover in the limit of vanishing equivariant parameter, the partition function reduces to the exponential of the on-shell value of the exact twisted superpotential computed in \cite{Witten1993}.  As discussed in \cite{Dimofte2010}, the open refined topological string partition functions for the toric diagrams such as Figures \ref{1surface}, \ref{2surface} can reproduce the K-theoretic lift of two dimensional $\cN=(2,2)$ partition function, we therefore expect it can also be expressed in terms of the K-theoretic lift of the deformed twisted superpotential for Theory II.
Combining this with the fact that closed refined topological string amplitudes correspond respectively to the instanton partition of Theory I, we can readily interpret the exact correspondence between them as a manifestation of geometric transition in refined topological string.

In the next section we will perform explicit refined topological string computations for the simplest non-trivial example to illustrate various points discussed in this section.

\subsubsection*{Single Stack of Surface Operators : $\vec{n}=(0,0, \dots n_s, \dots 0, 0)$}
\paragraph{}
\begin{figure}
\centering
\includegraphics[width=160mm]{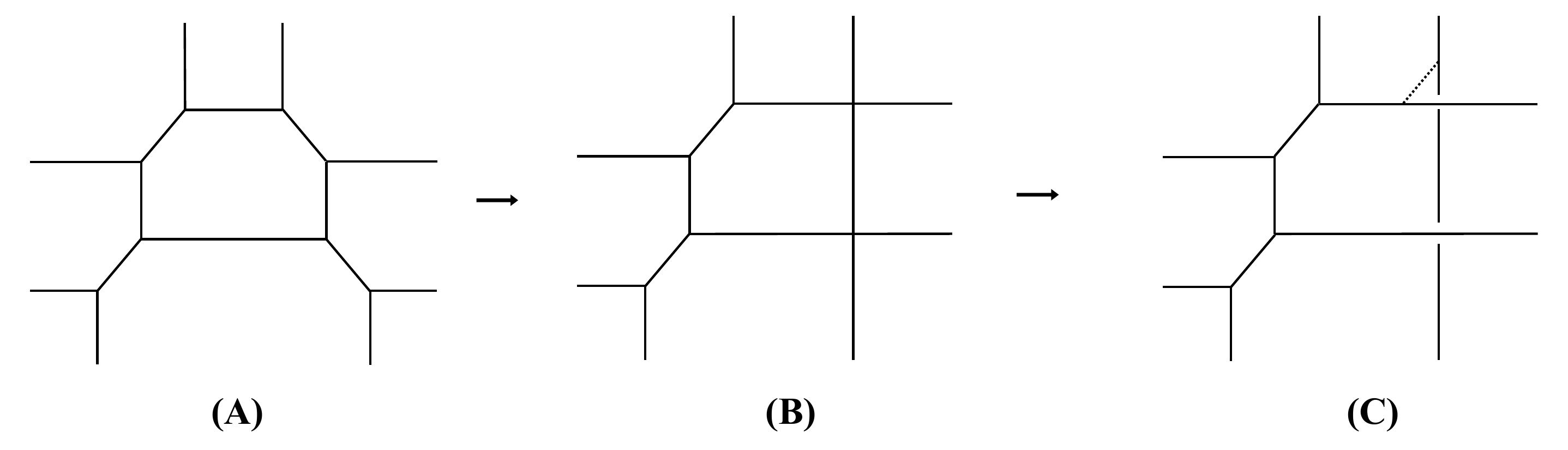}
\caption{Geometric Transition with Single Toric Brane Insertion}
\label{1surfaceGeo}
\end{figure}
Here as a consistency check, we consider the single stack of toric brane insertion with only $n_s \neq 0$ for the Abelian factor $U(1)_s$, they are labeled by a Young diagram with $n_s$ columns, this should be regarded as a slight generalization of the refined geometric transition involving single toric brane considered in \cite{Dimofte2010, Taki2010}.
We first notice that $\cZ_{\rm OV}$ now reduces precisely to unity, this is consistent with our interpretation of this configuration as single stack of $n_s$ toric branes insertion with holonomy given by Young diagram $Y_{s}$ in the refined topological string partition function. Here there are no additional contributions from M2 branes stretching between different stacks of toric branes to generate such factor.
The expression (\ref{ZinstTop}) now reduces to the following:
\begin{equation}
\Zinst(Q; q,t)
=\sum_{\{Y_s\}}  q^{\frac{1}{2}||Y_s ||^2} \tilde{\Lambda}^{2|Y_s |} P_{Y_s^{T}}(q^{\rho}, t^{n_s} q^{\rho}; t, q) \prod_{k=1}^{N_c}\prod_{i=1}^{n_s}\prod_{j=1}^{Y_{si}}\frac{\left(1-\frac{1}{v}\frac{Q_k^-}{ Q_s}q^{j-Y_{si}}t^{i-1}\right)}{\left(1-\frac{Q_k}{Q_s}q^{j-Y_{si}-1}t^{i-Y_{kj}^{\rm T}}\right)},
\label{trZinst2A}
\end{equation}
where $Y_{k}=0, k\neq s$. 
In \cite{Kozcaz2010}, a refined open topological vertex computation was sketched out for a single stack of $n_s$ toric branes with holonomy labeled by Young diagram $Y_s$,   inserted in the so-called $\tilde{T}_{N_c}$ geometry which corresponds to ``half'' of the toric geometry engineering the $SU(N_c)$, $N_f=2N_c$ gauge theory or ``strip'' geometry discussed earlier.    
The resultant open topological string amplitude was proposed to be given in terms of the so-called {\it $qt$-deformed hypergeometric functions} defined in \cite{Kaneko} (see equation (5.20) of \cite{Kozcaz2010} for explicit definition), 
and the authors of \cite{Kozcaz2010} verified this proposal by considering the unrefined limit $q=t$ of topological string computations.
Here we immediately observe that in our case, by expanding out the denominators in the truncated five dimensional Nekrasov partition function (\ref{trZinst2A}) and matching the parameters, we can rewrite it  in terms of the $qt$-deformed hypergeometric function defined in \cite{Kozcaz2010} as {}\footnote{Note that in \cite{Kozcaz2010}, the convention used was $(q,t)=(e^{\epsilon_1}, e^{-\epsilon_2})$, this requires us to change conventions accordingly.}:
\begin{eqnarray}
&&\Zinst(Q; q,t)={}_{N_c}\Phi_{N_c-1}^{(q^{-1}, t^{-1})}(\{a_\alpha \}_{N_c} ; \{b_\beta\}_{N_c-1} ; \{z_{\gamma}\}_{n_s})\nonumber\\
&&a_\alpha =\frac{1}{v}\frac{Q_\alpha^-}{ Q_s} , \quad \alpha=1,\dots, N_c, \quad b_{\beta}=\frac{1}{v^2}\frac{Q_\beta}{Q_s}, \quad \beta=1,\dots \neq s,\dots N_c, \nonumber\\
&& z_{\gamma}=\tilde{\Lambda}^2 t^{\gamma - 1}, \quad \gamma=1,\dots, n_s,
\end{eqnarray}
This matches with the expected expression in \cite{Kozcaz2010} for general $(q, t)$.
If we further set $n_s=1$, the MacDonald polynomial reduces to Schur function and the resulting expression reduces for the refined topological string amplitude with single toric brane insertion obtained from the explicit computation  in \cite{Taki2010}.
The observation here provides further confirmation to the refined geometric transition interpretation  proposed here.
\begin{figure}
\centering
\includegraphics[width=160mm]{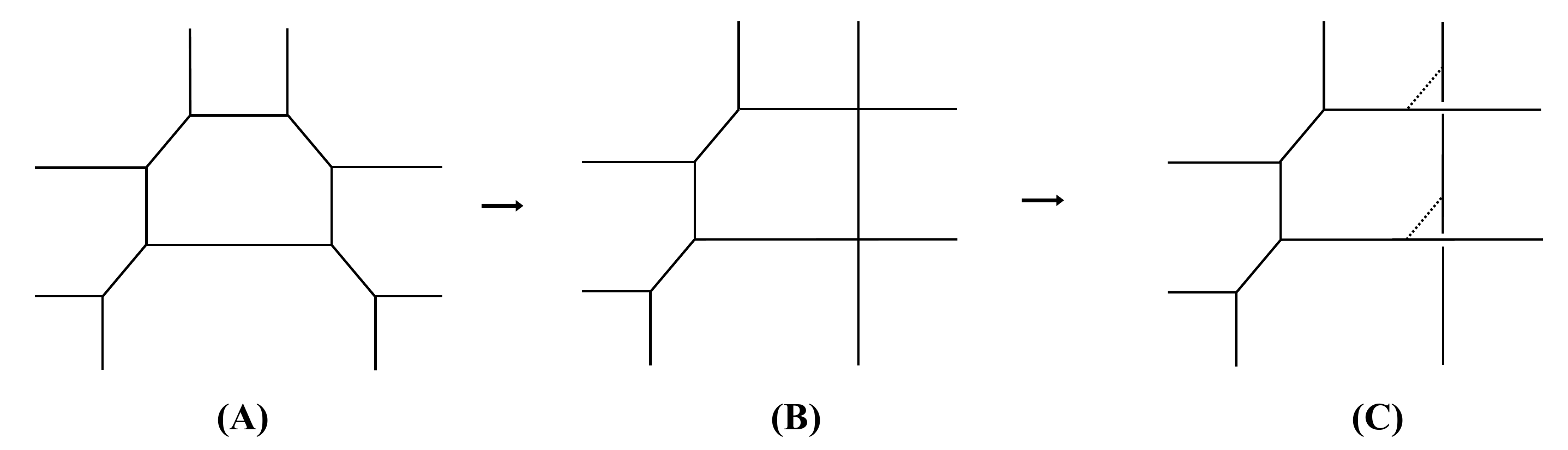}
\caption{Geometric Transition with Two Toric Brane Insertions}
\label{2surfaceGeo}
\end{figure}
\subsection*{Multiple of Single Surface Operators :  $\vec{n}=(1, 1,\dots ,1 ,1)$ }
\paragraph{}
Another interesting case is when we have $\vec{n}=(1,1,\dots, 1, 1)$,  such that the truncated instanton partition function is now labeled by a set of Young diagrams of single column with lengths $(L_1, \dots L_k, \dots, L_{N_c})$.  It is important to realize that the factor $\cZ_{\rm OV}$ is no longer unity, but we have:
\begin{equation}\label{ZOV2}
\cZ_{\rm OV}=\prod_{l>k}^{N_c} \frac{\prod_{j=1}^{L_l}(1-\frac{Q_l}{Q_k}t q^{j-1})\prod_{j=1}^{L_k}(1-\frac{Q_k}{Q_l}t q^{j-1})}{\prod_{j=1}^{L_l}(1-\frac{Q_l}{Q_k}q^{j-L_k-1})\prod_{j=1}^{L_k}(1-\frac{Q_l}{Q_k}t q^{L_l-j})},
\end{equation}
This factor now comes from M2 brane modes stretching between different single toric branes.
Combining with other factors, the truncated instanton partition function now reduces to
\begin{eqnarray}
\Zinst(Q;q,t)&=&\sum_{\{\vec{L}\}} \prod_{l,k=1}^{N_c} \tilde{\Lambda}^{2 L_k}\frac{\left(v\frac{Q_k}{Q_l^-} ; q\right)_{L_k}}{\left(q\frac{Q_k}{Q_l} ; q\right)_{L_k}} \left(q\frac{Q_k}{Q_l}; q\right)_{L_k-L_l}\nonumber\\
&=&\sum_{\{\vec{L}\}}\prod_{l,k=1}^{N_c}\tilde{\Lambda}^{2L_k}
\prod_{r=1}^{L_k}\frac{(1-v\frac{Q_k}{Q_l^-}q^{r-1})}{(1-\frac{Q_l}{Q_k}q^r)}\prod^{L_k-L_l}_{s=1}\left(1-\frac{Q_l}{Q_k } q^s\right) \label{Zinst1B}.
\end{eqnarray} 
We shall discuss how to reproduce this expression for $N_c=2$ case from direct open refined topological string computation in the next section.


\section{Explicit Example of Refined Geometric Transition}
\paragraph{}
In this section, we calculate the closed and open refined topological string partition functions for the configurations given in Figures \ref{fullgeometry}, \ref{1surface} and \ref{2surface}  as an explicit illustrations of the refined geometric transition discussed in the previous section.
Here in the closed topological string computation we shall consider $N_c=2$ and $N_f=4$ , for general $N_c$ the corresponding computation has been performed in \cite{Taki2007, Awata2008}. We review this simplest case to set the notations, and this is also sufficient to 
illustrate the connection between the geometric transition and exact correspondence between the gauge theories we proposed earlier. The discussion here can readily be generalized to higher rank cases. In  Appendices A and B we specify the refined topological vertex used in our computation and also derive the superpotential from the WKB limit of open refined topological string partition function.

\subsection*{The Closed Refined Topological String}
\paragraph{}
First we consider the toric geometry given in Figure \ref{fullgeometry} and divide it into left and right strips\footnote{Notice that the identifications between $(q, t)$ and equivariant parameters used in \cite{Taki2007} was $(q,t)=(e^{-R\epsilon_1}, e^{R\epsilon_2})$,  in other words related to the identification used in (\ref{Notations}) by $(q, t) \to (t^{-1}, q^{-1})$ transformation. However as the final expression can be written in terms of the function $\cN_{Y_{l} Y_{k}}(Q; q, t)$ given in (\ref{DefcN}), which enjoys the identity $\prod_{l,k=1}^{N_c}\cN_{Y_l Y_k} (Q; t^{-1}, q^{-1}) = \prod_{l,k=1}^{N_c}\cN_{Y_l^{\rm T} Y_k^{\rm T}} (Q; t^{-1}, q^{-1})$, we merely need to take into account of the transpose of Young diagrams and exchange the columns and rows. We have also restored the radius $R$ of $S^1$ in this section.}.
We now apply the refined topological vertex $C_{\lambda\mu\nu}$ of \cite{IKV}, which is reviewed here in Appendix A, to the left strip geometry given in Figure \ref{halfgeometry}. Assigning the Young diagrams $Y_{1,2}$ to the horizontal legs, we can write down the corresponding contribution as
\bea
&&Z^{L}_{Y_1Y_2}(t,q;  \qmo, \qmt, \qfo) = \nn  \\
&& \sum_{R_i} (- \qmo)^{\aro} (-\qfo)^{\art} (-\qmt)^{\arth} C_{R_1 0 Y_1} (t, q) C_{\rot R_2 0} (q,t) C_{R_3 \rtt Y_2} (t, q) C_{R_3^T 0 0} (q, t)  \nonumber \\
&&= Z (Y_1, Y_2) \sum_{R_i} s_{\rot} (\qmo \qfo t^{-\rho} q^{-Y_1})  \nn 
 \sum_{\eta} s_{R_1 /\ \eta} 
 \left(-({1 /\  \qfo}) q^{-\rho} \right ) s_{R_2 /\ \eta}  (-  t^{-\rho} \qfo \sqrt{q /\ t} )  \nn \\
&& \times \sum_{\gamma} s_{\rtt /\ \gamma} (\sqrt{t /\ q} \, t^{-Y_2^T} q^{-\rho})
s_{\rtth /\ \gamma} (t^{-\rho} q^{-Y_2}) s_{R_3} (- \qmt q^{-\rho})\label{ZL1}
\eea
where
\be 
Z(Y_1, Y_2) = q^{{||Y_1||^2 \over 2} +{ ||Y_2||^2 \over 2}} {\tilde Z}_{Y_1} (t, q) {\tilde Z}_{Y_2} (t, q).
\ee
Using the Schur function summation identities the summation over intermediate Young diagram $\{R_3\}$ in (\ref{ZL1}) can be evaluated as
\bea
&&Z^{L}_{Y_1Y_2}(t,q;  \qmo, \qmt, \qfo) = \nn  \\
&&= Z (Y_1, Y_2) \sum_{R_{1,2}} s_{\rot} (\qmo \qfo t^{-\rho} q^{-Y_1})  \nn 
 \sum_{\eta} s_{R_1 /\ \eta} 
 \left(-({1 /\  \qfo}) q^{-\rho} \right ) s_{R_2 /\ \eta}  (-  t^{-\rho} \qfo \sqrt{q /\ t} )  \nn \\
&& \times \sum_{\gamma} s_{\rtt /\ \gamma} (\sqrt{t /\ q} \, t^{-Y_2^T} q^{-\rho})
s_{\gamma} (- \qmt q^{-\rho}) \prod_{i, j} (1 - \qmt q^{i - \half - Y_{2j}} t^{j- \half})
\eea
and proceeding similarly with summations over $\{R_{1,2}\}$ we arrive at
\bea
&&Z^{L}_{Y_1Y_2}(t,q;  \qmo, \qmt, \qfo) = Z(Y_1, Y_2)  \times \nn \\
&\times& \prod_{i,j =1}^{\infty} {(1 - \qmo t^{i - \half} q^{-Y_{1i} + j - \half})  (1- \qmo \qfo \qmt 
 t^{j- \half} q^{-Y_{1j} + i - \half}) \over (1 - \qmo \qfo t^{- Y_{2i}^{T}  + j}  q^{-Y_{1j} + i -1})} \nn \\
 &\times& \prod_{i,j =1}^{\infty} { (1 - \qfo q^{j - \half} t^{-Y_{2j}^T + i - \half}) (1 - \qmt t^{i - \half} q^{-Y_{2i} + j - \half}) \over (1 - \qfo \qmt t^{i-1} q^j)}.
\eea
This agrees completely with (3.2) of \cite{Taki2007}.

For the full geometry we glue the right hand part as in Figure \ref{fullgeometry}.  
The right partition function can be given similarly as:
\bea
&&Z^{R}_{Y_1, Y_2} (t, q;  \hqmo, \hqmt, \hqfo) =   t^{{||Y_1^T||^2 \over 2} +{ ||Y_2^T||^2 \over 2}} {\tilde Z}_{Y_1^T} (q, t) {\tilde Z}_{Y_2^T} (q, t) \times \nn \\
&\times& \prod_{i,j =1}^{\infty} {(1 - \hqmo t^{i - \half} q^{-Y_{1i} + j - \half})  (1- \hqmo \hqfo \hqmt 
 t^{j- \half} q^{-Y_{1j} + i - \half}) \over (1 - \hqmo \hqfo t^{- Y_{2i}^{T}  + j-1}  q^{-Y_{1j} + i})} \nn \\
 &\times& \prod_{i,j =1}^{\infty} { (1 - \hqfo q^{j - \half} t^{-Y_{2j}^T + i - \half}) (1 - \hqmt t^{i - \half} q^{-Y_{2i} + j - \half}) \over (1 - \hqfo \hqmt t^{i} q^{j-1})}.
\eea
The full partition function is therefore
\be
Z_{{\rm closed}} = \sum_{Y_1, Y_2} Q^{|Y_1| + |Y_2|}  \frac{Z^{L}_{Y_1Y_2}(t,q;  \qmo, \qmt, \qfo) 
Z^{R}_{Y_1, Y_2} (t, q;  \hqmo, \hqmt, \hqfo)}{Z^{L}_{0,0}(t,q;  \qmo, \qmt, \qfo) Z^{R}_{0,0} (t, q;  \hqmo, \hqmt, \hqfo)}  \label{fullform1}
\ee
where we normalized it with respect to the partition function with empty Young diagrams in order to match with the gauge theory instanton partition function momentarily. Furthermore, we use the following conversion between geometric moduli and gauge theory parameters: 
\bea
Q_{12} &=& \qmo \qfo =  e^{-2 R a} \nn \\
\qmo &=& e^{R (m_1 - a)} t^{\half} q^{-\half} \nn \\
\qmt  &=& e^{R(m_2  + a)} t^{\half} q^{-\half} \nn \\
\hat{Q}_{12} &=& \hqmo \hqfo = e^{-2 Ra } \nn \\
\hqmo &=& e^{R({\hat m_1} -a)}  t^{\half} q^{-\half} \nn  \\
\hqmt &=& e^{R({\hat m_2} + a)} t^{\half} q^{-\half} \nn
\eea
Here $a=a_1=-a_2$ is the Coulomb branch parameter for $SU(2)$ gauge group and we also introduced the masses of the fundamentals $m_{1,2}$ and ${\hat m}_{1,2}$.  
After making such identification with the gauge theory parameters, we can recast the closed refined topological string amplitude as the instanton partition function for the five dimensional ${\mathcal{N}}=1$ $SU(2)$ gauge theory with $N_f=4$ flavors, i. e. the simplest case of Theory I.
For the later use, here we also recast the left strip partition function in terms of gauge theory parameters as:
\bea
&&Z^{L}_{Y_1Y_2}(t,q;  a, m_1, m_2) = q^{{||Y_1||^2 \over 2} +{ ||Y_2||^2 \over 2}} {\tilde Z}_{Y_1} (t, q) {\tilde Z}_{Y_2} (t, q) \times \nn \\
&\times& \prod_{i,j =1}^{\infty} {(1 - e^{R(m_1 - a)}  t^{i - 1} q^{-Y_{1i} + j})  (1-  e^{R (m_2 -a)} t^{j-1}
  q^{-Y_{1j} + i }) \over (1 - e^{-2 R a}  t^{- Y_{2i}^{T}  + j}  q^{-Y_{1j} + i -1})} \nn \\
 &\times& \prod_{i,j =1}^{\infty} { (1 - e^{-R (m_1 + a)}  q^{j - 1} t^{-Y_{2j}^T + i}) (1 -e^{R (m_2 + a)} t^{i - 1} q^{-Y_{2i} + j}) \over (1 - e^{R(m_2 - m_1)}  t^{i-1} q^j)}
\eea
and similarly we can deduce that the contribution from right strip is given by:
\bea\label{ClosedAmp}
&&Z^{R}_{Y_1, Y_2} (t, q;  a, \hmo, \hmt) =   t^{{||Y_1^T||^2 \over 2} +{ ||Y_2^T||^2 \over 2}} {\tilde Z}_{Y_1^T} (q, t) {\tilde Z}_{Y_2^T} (q, t) \times \nn \\
&\times& \prod_{i,j =1}^{\infty} {(1- e^{R(\hmo -a)}  t^{i-1 } q^{-Y_{1i} + j})  (1-  e^{R(\hmt -a)} 
 t^{j-1}  q^{-Y_{1j} + i}) \over (1 - e^{-2 Ra} t^{- Y_{2i}^{T}  + j-1}  q^{-Y_{1j} + i})} \nn \\
 &\times& \prod_{i,j =1}^{\infty} { (1 - e^{-R(\hmo +a) } q^{j-1} t^{-Y_{2j}^T + i}) (1 - e^{R(\hmt +a)} t^{i-1} q^{-Y_{2i} + j}) \over (1 - e^{R(\hmt - \hmo)}  t^{i} q^{j-1})}.
\eea
In the full partition function the K\"ahler parameter of the base $\bP^1$ is identified in terms of the five dimensional gauge theory coupling $\tau$ as $Q = e^{2 \pi i \tau}$.
The closed refined topological string can then be recast into the five dimensional gauge theory instanton partition function with $N_f=2N_c$, using the formulae provided in \cite{Awata2008}, see also \cite{Taki2007, Awata2008} for higher rank generalization.
\subsection*{Open string partition function}
\paragraph{}
Let us now move to the open refined topological string amplitude, and first examine the simplest case of a single brane placed on one internal leg of the toric diagram, as shown explicitly in Figure \ref{1surface}. 
In this case $Y_2=0$, and the MacDonald polynomial corresponding to the gauge holonomy reduces to a single power of toric brane position and the expression simplifies to
\be
Z_{\rm open}= \sum_{[Y_1]^{1}} z^{|Y_1|}  Z_{Y_1 0}^L (t,q; \qmo, \qmt, \qfo)
\ee
where the sum becomes restricted to single column partitions. 
Since we want to compare this with the normalized closed string partition function,
let us also normalize the open string partition function as
\be
Z_{\rm open}= \sum_{[Y_1]^{1}} z^{|Y_1|} { Z_{Y_1 0}^L (t,q;
\qmo, \qmt, \qfo) \over Z_{00}^L(t,q;
\qmo, \qmt, \qfo)} \label{openpartnormalized}.
\ee
The insertion of a toric brane into left strip geometry precisely corresponds to the D-brane set up of Theory II via geometric engineering \cite{Dimofte2010}.
As pointed out in \cite{Dimofte2010, Bonelli2011} that refined topological string amplitude for such strip geometry with toric brane insertion can yield the K-theoretic lift of various two dimensional $\cN=(2,2)$ vortex partition functions.

We compare this directly with the general form (\ref{fullform1}) 
of the closed string partition function with the degenerate K\"ahler moduli 
$a =  -{\hat m}_1= {\hat m}_2  +  \epsilon_1$ substituted.
Let us now examine $Z^{R}_{Y_1, Y_2}/Z^{R}_{0, 0}$ in the resultant expression, first we can drop the summation over empty $Y_2$,
and the partition function simplifies into:
\be
Z_{{\rm closed}} = \sum_{Y_1} e^{2\pi i \tau |Y_1|}  \frac{Z^{L}_{Y_10}(t,q;  \qmo, \qmt, \qfo)
Z^{R}_{Y_1, 0} (t, q;  \hqmo, \hqmt, \hqfo)}{Z^{L}_{0,0}(t,q;  \qmo, \qmt, \qfo) Z^{R}_{0,0} (t, q;  \hqmo, \hqmt, \hqfo)}  \label{fullform2}
\ee
where the right strip contribution above reduces to
\be
Z^{R}_{Y_1, 0}/Z^{R}_{0, 0} = {t^{{||Y_1^T||^2 \over 2}} {\tilde Z}_{Y_1^T}(q,t) 
\prod_{i,j=1}^{\infty} (1- t^{j-1} q^{-Y_{1j} + i +1} ) \over
\prod_{i,j=1}^{\infty} (1- t^{j-1} q^{ i +1} )} ,
\ee
for a given single column partition of length $L$, $Y_1= [1]_L$ 
\be
Z^{R}_{Y_1, 0}/Z^{R}_{0, 0} = {t^{L^2 \over 2}} {\tilde Z}_{Y_1^T}(q,t) \prod_{i=1}^L (1 - t^{i-1} q). 
\ee
Moreover for a single column Young diagram the framing factor $\tilde{Z}_{Y_1^{\rm T}}$ becomes:
\be
{\tilde Z}_{Y_1^{\rm T}}(q,t) = \frac{1}{\prod_{i=1}^L (1 - t^{i-1} q)}.
\ee 
This allows us to reduce the contribution from the right strip to just a framing factor $t^{L^2/2}$.
Thus the closed string partition function equals to the open string
partition function up to a framing factor
\be
Z_{{\rm closed}} = \sum_{Y_1^1} e^{2\pi i\tau |Y_1|} t^{{||Y_1^T||^2 \over 2}} \frac{Z^{L}_{Y_10}(t,q;  
\qmo, \qmt, \qfo)}{Z^{L}_{0,0}(t,q;  \qmo, \qmt, \qfo)}.
\ee
Following the derivation outlined in \cite{Taki2010} one can also easily find the effective superpotential and the associated Gaiotto curve from the open string partition function. This instructive computation is done in our Appendix B.

\subsection*{Two brane insertions and Geometric Transition}
\paragraph{}
Finally let us consider the insertion of two single toric branes on the two distinct internal legs of the toric diagram, as shown in Figure \ref{2surface}, 
the gauge holonomies on them are labeled by Young diagrams $Y_1$ and $Y_2$ respectively.
Applying the rule of refined topological vertex, the normalized partition function for this configuration can be written as:
\be\label{Zopen2}
Z_{\rm open} = \sum_{Y_1,Y_2} {\rm Tr}_{Y_1} U\, {\rm Tr}_{Y_2} V {Z^L_{Y_1 Y_2} (q,t) \over
Z^L_{00}(q,t) } \times \cZ_{\rm OV},
\ee
where ${\rm Tr}_{Y_1}U$ and ${\rm Tr}_{Y_2}V$ are gauge holonomies and the eigenvalue for the matrices $U$ and $V$ label the position of the toric branes,
$Z^L_{Y1 Y_2}(q,t)/Z^L_{00}(q,t)$ is the normalized right strip contribution and $\cZ_{\rm OV}$ is the refined Ooguri-Vafa factor.
Denoting the positions of the two toric branes as $z_{1,2}$ we obtain
\be
Z_{\rm open} = \sum_{[Y_1]^1, [Y_2]^1} z_{1}^{|Y_1|} z_{2}^{|Y_2|}  {Z^L_{Y_1 Y_2} (q,t) \over
Z^L_{00}(q,t) } \times \cZ_{\rm OV}.
\ee
Performing the summation for the single column Young diagrams of length $L_1$ and $L_2$ this expression 
can be written as
\bea
Z_{\rm open} &=& \sum_{L_1,L_2=0}^{\infty} z_1^{L_1} z_2^{L_2} t^{{L_1+L_2 \over 2}} \times \nn \\
&&\prod_{i=1}^{L_1} {(1 - e^{R(m_1 -a)} q^{i-1}) 
(1 - e^{R(m_2 -a)} q^{i-1}) \over (1-q^i)}  \prod_{k=1}^{L_1} {(1 - e^{-2 aR}
q^{-L_2 + k}) \over (1 - e^{-2 aR} q^{-L_2 + k}/t) (1 - e^{-2aR} q^k) }  \times
\nn \\
&&\prod_{j=1}^{L_2} {(1 - e^{-R(m_1+a)} q^{-j+1}) (1 - e^{R(m_2 +a)} q^{j-1})
\over (1 - e^{-2aR} q^{-j +1}) (1 - q^j)} \times \cZ_{\rm OV},  
\eea
when $L_1 \rightarrow 0$ we get back the previous single brane open amplitude.

Here we can generalize our study of geometric transition for single toric brane insertion directly to this case, for the closed refined topological string amplitude (\ref{fullform1}) we impose the following double geometric transition condition: 
\be 
a = -\hmo - \epsilon_1 = \hmt + \epsilon_1.
\ee
In such a limit the left strip contribution ${Z^L_{Y_1 Y_2} (q,t) \over
Z^L_{00}(q,t) }$ remains intact  in both the refined open (\ref{Zopen2}) and closed (\ref{fullform1}) topological string partition functions,
while as discussed in the previous section, via geometric transition we expect the right strip contribution ${Z^R_{Y_1 Y_2} / Z^R_{00}}$ in (\ref{fullform1}) to match precisely to the remaining part of the open string amplitude. 

Substituting the degenerate parameters from the double geometric transition directly, we find that both $Y_1$ and $Y_2$ are now restricted to single column partitions. \footnote{Note that we could have taken instead
\be
a = -\hmo - \epsilon_2 = \hmt + \epsilon_2
\ee
which would restrict $Y_1$ and $Y_2$ both single column diagrams. So changing $\epsilon_1$ to $\epsilon_2$ does a transposition on the tableaux .}  We obtain 
\begin{eqnarray}
&&{Z^R_{Y_1 Y_2} (q, t) \over  Z^R_{00} (q, t) }= \left({-\sqrt{q} \over t}\right)^{L_2} t^{L_1^2 \over 2} 
\prod_{i=1}^{\infty} ( 1 - e^{-2 Ra}  q^{i-1} t) \prod_{j=1}^{L_2} ( 1 - e^{2 Ra} q^{j-1}t ) \nn \\
&& \times \prod_{i=1}^{L_1} {1 \over(1-  e^{-2 R a} q^{-L_2 + i-1} )} \prod_{i=L_1+1}^{\infty} {1 \over 
(1 - e^{-2 Ra} t q^{-L_2 + i-1})}
\end{eqnarray}
Comparing this expression to to $\cZ_{\rm OV}$,  from Figure \ref{2surface} we can deduce it consists of M2 brane modes stretching between the two toric branes themselves and from each toric brane to the background flavor branes.  We can then directly use the results for the refined Ooguri-Vafa factor in \cite{AS2012} and deduce that it precisely matches with the right strip contribution, which completes the refined geometric transition.

\section{Future Directions}
\paragraph{}
In this note, we interpreted the exact correspondence proposed in \cite{DHL, CDHL}, originally arising from connecting supersymmetric gauge theories sharing the same integrable structure as a realization of geometric transition in refined topological string. Here we would like to wrap up discussing  some interesting future directions.

Although we have only looked at the specific examples, the refined geometric transition analysis presented here in fact gives a general prescription for relating two distinct supersymmetric gauge theories, one with and the other without the vortices/surface operator insertions. Consider starting instead with a 4d linear quiver theory \footnote{For example see section 3 of \cite{CDHL} and also section 3.2 of \cite{Chen2012A} for elliptic generalization,  the simplest case of $SU(2)\times SU(2)$ was considered in \cite{Taki2010}.}, and arrange the Coulomb parameters and bi-fundamental masses such that only partial gauge nodes in the quiver are on their baryonic Higgs roots. After the refined geometric transition we end up with a coupled 2d-4d system involving both the dynamics of vortices and residual 4d gauge theory. By decoupling the vortex dynamics, this procedure also predicts the equivariant partition function of gauge theories with surface operator insertions, it would be interesting to verify these predictions from the direct field theoretic localization computations. 
Moreover, to connect with the celebrated conjecture by Alday, Gaiotto and Tachikawa \cite{AGT}, it would also be interesting to consider how these partition functions with general surface operator insertions and their non-trivial mutual interactions can be reproduced by the dual conformal field theories with appropriate vertex operator insertions \cite{AGGTV, AT2010}. 

Another direction to consider is that through the presence of vortices, a 4d $\cN=2$ supersymmetric gauge theory can have two possible 2d dual theories, one is the supersymmetric vortex world volume theory extensively discussed here and the other one is the non-supersymmetric conformal field theory proposed in \cite{AGT, AGGTV}, and so it is natural to consider their connections.   A clue is that these two 2d field theories can both have connections with different integrable systems, such as the quantum spin chain from the quantum vacua of vortex theory considered here, and Gaudin model from the KZ equation in WZW conformal field theories \cite{Feigin1994, Teschner2010}. It turns out that these two distinct integrable models are related through what is known as bi-spectral duality in integrable system literature, such that the spectral curves and the degrees of freedom of two integrable system can be identified.  Recently this duality has been applied in \cite{Mironov2012A, Mironov2012B, BCGK} to verify the AGT conjecture in the NS limit. It would also be important to consider in the most general case where both equivariant parameter $\epsilon_{1,2}$ are non-vanishing, and how these 2d field theories can be related.

Finally, in the presence of world volume equivariant parameter, the three dimensional supersymmetric vortex theory on $R^2\times S^1$ belongs to the class considered in \cite{Dimofte2010, Dimofte2011A, Dimofte2011B, Terashima2011} \footnote{More precisely, the equivariant partition function on $R^2\times S^1$ as the building holomorphic block for 3d partition functions and index under the factorization as explained in \cite{Pasquetti2011, Beem2012}.}, where a 3d/3d duality with Chern-Simon theory was proposed. Exact quantities such as the index, the squashed three sphere partition function, and the Wilson loop expectation value in 3d supersymmetric gauge theories can be mapped to the corresponding quantities in the dual 3d Chern-Simons theory on various three manifolds. Given the exact 3d/5d correspondence proposed here, it is natural to ask whether there can be concrete connections between the 5d supersymmetric gauge theory and the 3d Chern-Simons theory, possibly through a chain of dualities in the refined topological theory. We hope to return to these directions in the near future.

\acknowledgments
We would like to thank Nick Dorey for initial collaboration on this work, and for various insightful questions and discussions. HYC would also like to thank the University of Cincinnati and University of Kentucky for the generous financial supports where most of this work was carried out, he would also like to thank Po-Shen Hsin for helping to prepare the figures. His research is also supported in part by National Science Council and Center for Theoretical Sciences at National Taiwan University.

\section*{Appendix A: The refined topological vertex}\label{AppA}
\paragraph{}
Here we use the representation of the refined vertex presented in \cite{IKV}  and also used in\cite{Taki2010}
\bea
&&C_{R_1 R_2 R_3} (t, q) \\
&&=  \left( q \over t \right)^{ (||R_2||^2 + ||R_3||^2) \over 2} t^{\kappa_{R_2} \over 2}
P_{R_3^{T}}(t^{-\rho}; q, t) \sum_{ \eta}   \left( q \over t \right)^{(|\eta| + |R_1| - |R_2|) \over 2}  
s_{R_1^T /\ \eta} (t^{-\rho} q^{-R_3}) s_{R_2 /\ \eta} ( t^{- R_3^T} q^{-\rho}) \nonumber
\eea
where  $q= e^{ -\epsilon_1}$,  $t=e^{\epsilon_2}$, $\kappa_Y=\sum_{(i,j) \in Y} (j-i)=||Y||^2-||Y^{\rm T}||^2$ and $P_{R;  q, t}$ is the Macdonald function
\be
P_{R^T}(t^{-\rho}; q, t)  = t^{||R||^2 \over 2} {\tilde Z}_R(t, q) \quad {\tilde Z}_R(t, q) = \prod_{(i,j) \in R} 
\left( 1 - t ^{R_j^T - i + 1} q ^{R_i -j} \right )^{-1}
\ee

In the A-model limit $q=t= e^{ -g_s}$ this expression reduces to the topological vertex representation
\be
C_{R_1 R_2 R_3} (q) = q^{\kappa_{R_2} \over 2} s_{R_3^T} (q^{-\rho})
 \sum_{\eta} s_{R_1^T /\ \eta} ( q^{-R_3 - \rho})  s_{R_2 /\ \eta}  ( q^{-R_3^T - \rho})
\ee

\section*{Appendix B:  Superpotenial from the open string partition function}\label{AppB}
\paragraph{}
Here we rewrite the single brane open string partition function in its simplest form. This is the explicit form of the normalized  open string partition function appearing in \cite{Taki2010}  most suitable to extract the differential equation it obeys. We start with
\be
Z (q, t)= \sum_{[Y_1]^{1}} z^{|Y_1|} { Z_{Y_1 0}^L (q,t;
\qmo, \qmt, \qfo) \over Z_{00}^L(q,t;
\qmo, \qmt, \qfo)}  \label{openpartnormalized}
\ee
Using that $Y_1$ are single row partitions, we can sum over all such partitions of a given length $L$.
Rewriting as an exponential,  and performing the sums we obtain
\bea
Z (q, t) &=& \sum_{L=0}^{\infty} z^L t^{L \over 2}
   \prod_{i=1}^{L} {( 1 - e^{R(m_1 -a)}  q^{i-1}) ( 1 - e^{R(m_2 -a)} q^{i-1}) \over (1 - q^{i}) ( 1 - e^{-2 Ra} {q^i / t}) } \nn \\
  &=& \sum_{L=0}^{\infty} z^L Z^{(L)}
\eea
Here we get perfect agreement with \cite{Taki2010} (2.18), apart from the framing factor 
$t^{L/2}$, which we can absorb in the rescaling of $z \rightarrow
 \sqrt{t} z$.\footnote{In \cite{Taki2010} the 
framing factor disappears by the modification of the vertex rules with
$f_{R}(t, q) = (-1)^{|R|} t^{||R^{T}||^2/2} q^{-||R||^2/2}$, as described in
\cite{Taki2007}.}
 
Defining $z = e^{-u}$ we get
\be
e^{-Lu} Z^{(L)} = 
{ (1 - q^{L-1} e^{R(m_1 -a)}) (1 -  q^{L-1} 
e^{R(m_2 -a)}) \over (1 - q^L)  (1 - e^{-2 Ra} {q^L / t}) } z 
e^{-(L-1) u} Z^{(L-1)} 
\ee
This leads to the differential equation
\be
\left[(1 - e^{-2 Ra} q^{-\partial_u} /t)  (1 - q^{-\partial_u}) - z  ( 1-  e^{R(m_1 -a)} q^{- \partial_u} )
( 1-  e^{R(m_2 -a)} q^{- \partial_u} ) \right] Z^{open} =0 
\ee
It is easy to read off the geometry from here. We note that the Nekrasov-Shatashvili
limit of the full differential equation ($t \rightarrow 1$) is simply
\be
\left[(1 - e^{-2 Ra} q^{-\partial_u})  (1 - q^{-\partial_u}) - z  ( 1-  e^{R(m_1 -a)} q^{- \partial_u} )
( 1-  e^{R(m_2 -a)} q^{- \partial_u} ) \right] Z^{open} =0 \label{fivedeq}
\ee
as the t-dependence affects only in a single factor of the full differential 
equation.

Let us first examine the $R \to 0$ limit. In the field
theory limit $q=t$ this was already worked out in \cite{Taki2010}, where the Gaiotto 
curve and the superpotential was extracted. Here we keep
$$q = e^{- R \epsilon_1} \quad t =  e^{R \epsilon_2}.$$
Taking the $R\to 0$ limit,  we obtain for the partition function
\be
Z(\epsilon_1, \epsilon_2, t) = \sum_{L=0}^{\infty} z^L \prod_{i=1}^L
{(a - m_1 + (i-1) \epsilon_1) (a - m_2 + (i-1) \epsilon_1) \over
i \epsilon_1 (2a + i \epsilon_1 + \epsilon_2)} 
\ee
and for the differential equation
\be
\left[ -(2 a - \epsilon_1 \partial_u + \epsilon_2)
\epsilon_1  \partial_u - z ( (m_1 - a) + \epsilon_1 \partial_u) 
((m_2 - a) + \epsilon_1 \partial_u) \right] Z_{open} =0
\ee
or
\be
\left[ (z-1) \epsilon_1^2 \partial_u^2 + (2a  + \epsilon_2  +
z (m_1 -a) + z (m_2 -a)) \epsilon_1 \partial_u + z (m_1 -a) (m_2 -a) 
\right] Z_{open} =0
\ee
Taking the WKB limit 
\be
Z_{open} = e^{-{1 \over \epsilon_1} W(z) + \ldots}
\ee
we find almost the same differential equation for the superpotential 
as the field theory limit of \cite{Taki2010}
\be
(z-1) (\partial_u W(z))^2 - (2a  + \epsilon_2 +
z (m_1 -a) + z (m_2 -a)) \partial_u W(z) + z (m_1 -a) (m_2 -a)
=0
\ee
The difference is now in the $\epsilon_2$ dependence, which can be
absorbed as a shift in the parameters. 
Solving for $W(z)$ which can also be identified with the twisted  superpotential of the 2d gauged linear sigma model after change of parameters, 
we obtain 
\be
W(z) = \alpha_2 \log(z) + \alpha_3 \log(1 -z)  \pm 
\int^{z} { \sqrt{\alpha_1^2 (z'^2 -z') + \alpha_2^2 (1 -z') + \alpha_3^2 z'}
\over z' (1 -z')}  dz'
\ee
where
\bea
a- m_1 &=& - \alpha_1 + \alpha_2  -\alpha_3 \nn \\
-a - m_2 - \epsilon_2 &=&  \alpha_1 - \alpha_2  -\alpha_3 \nn \\
a + m_1 + \epsilon_2 &=& \alpha_1 + \alpha_2 + \alpha_3
\eea
or
\bea
\alpha_1 &=& \frac{m_1 - m_2}{2} \nn \\
\alpha_2 &=&  a + \frac{\epsilon_2}{2} \nn \\
\alpha_3 &=& \frac{m_1 + m_2}{2} + \frac{\epsilon_2}{2}
\eea
The shift can be absorbed in only one parameter, if we take the parametrization
$(\alpha_1, \alpha_3 -\alpha_2, \alpha_3)$ for example.
This also means that in the NS limit we obtain the same Gaiotto curve of 4d supersymmetric gauge theory, with a
shift in its parameters\footnote{
In the limit $\epsilon_2 \rightarrow 0$ or superpotential becomes
precisely that of (2.14) \cite{Taki2010}.}.

We can also get the mirror curve in the five dimensional geometry, after taking the NS limit we obtain
\be
H(e^p, e^u) + xy =0
\ee
where
$$p= - \epsilon_1 \partial u \quad [u, p] = \epsilon_1 = \hbar$$
and
\be
H(e^p, e^u) = (1- e^{-R(2a + \epsilon_2 + p)}) (1 -e^{-Rp})  - e^u ( 1- e^{R(m_1 -a -p)})
(1 - e^{R(m_2-a -p)}) 
\ee
If we compare with the field theory (q=t) limit, we find the same 
superpotential as in the usual field theory with the parameters shifted
as in four dimensions
\bea
a &\rightarrow&  a + \frac{\epsilon_2}{2} \nn \\ 
m_1  &\rightarrow& m_1 +  \frac{\epsilon_2}{2} \nn \\
m_2 &\rightarrow& m_2 +\frac{\epsilon_2}{2}
\eea
Hence the mirror curve should be given of that
of the usual field theory geometry, with shifted parameters.

\bibliographystyle{sort}

\begin{thebibliography}{sort}
\bibitem{DHL} 
  N.~Dorey, S.~Lee and T.~J.~Hollowood,
  JHEP {\bf 1110}, 077 (2011)
  [arXiv:1103.5726 [hep-th]].

\bibitem{CDHL} 
  H.~-Y.~Chen, N.~Dorey, T.~J.~Hollowood and S.~Lee,
  JHEP {\bf 1109}, 040 (2011)
  [arXiv:1104.3021 [hep-th]].

\bibitem{NSlimit} 
  N.~A.~Nekrasov and S.~L.~Shatashvili,
  ``Quantization of Integrable Systems and Four Dimensional Gauge Theories,''
  arXiv:0908.4052 [hep-th].

\bibitem{YangYang} 
  C.~-N.~Yang and C.~P.~Yang,
  J.\ Math.\ Phys.\  {\bf 10}, 1115 (1969).

\bibitem{Nekrasov2009A} 
  N.~A.~Nekrasov and S.~L.~Shatashvili,
  Nucl.\ Phys.\ Proc.\ Suppl.\  {\bf 192-193}, 91 (2009)
  [arXiv:0901.4744 [hep-th]].
    

\bibitem{Hanany1996} 
  A.~Hanany and E.~Witten,
  Nucl.\ Phys.\ B {\bf 492}, 152 (1997)
  [hep-th/9611230].

\bibitem{Hanany2004} 
  A.~Hanany and D.~Tong,
  JHEP {\bf 0404}, 066 (2004)
  [hep-th/0403158].

\bibitem{IKV} 
  A.~Iqbal, C.~Kozcaz and C.~Vafa,
  JHEP {\bf 0910}, 069 (2009)
  [hep-th/0701156].
  
\bibitem{Taki2007} 
  M.~Taki,
  JHEP {\bf 0803}, 048 (2008)
  [arXiv:0710.1776 [hep-th]].

\bibitem{Awata2008} 
  H.~Awata and H.~Kanno,
  Int.\ J.\ Mod.\ Phys.\ A {\bf 24}, 2253 (2009)
  [arXiv:0805.0191 [hep-th]].

\bibitem{Klemm1996} 
  A.~Klemm, W.~Lerche, P.~Mayr, C.~Vafa and N.~P.~Warner,
  Nucl.\ Phys.\ B {\bf 477}, 746 (1996)
  [hep-th/9604034].

\bibitem{Katz1996} 
  S.~H.~Katz, A.~Klemm and C.~Vafa,
  Nucl.\ Phys.\ B {\bf 497}, 173 (1997)
  [hep-th/9609239].

\bibitem{Karch1998} 
  A.~Karch, D.~Lust and D.~J.~Smith,
  Nucl.\ Phys.\ B {\bf 533}, 348 (1998)
  [hep-th/9803232].
  
\bibitem{Gopakumar1998} 
  R.~Gopakumar and C.~Vafa,
  Adv.\ Theor.\ Math.\ Phys.\  {\bf 3}, 1415 (1999)
  [hep-th/9811131].


\bibitem{Dimofte2010} 
  T.~Dimofte, S.~Gukov and L.~Hollands,
  Lett.\ Math.\ Phys.\  {\bf 98}, 225 (2011)
  [arXiv:1006.0977 [hep-th]].

\bibitem{Taki2010} 
  M.~Taki,
  JHEP {\bf 1107}, 047 (2011)
  [arXiv:1007.2524 [hep-th]].
  
\bibitem{AS2011} 
  M.~Aganagic and S.~Shakirov,
  arXiv:1105.5117 [hep-th].

\bibitem{AS2012} 
  M.~Aganagic and S.~Shakirov,
  arXiv:1210.2733 [hep-th].

   
\bibitem{Nekrasov2002} 
  N.~A.~Nekrasov,
  Adv.\ Theor.\ Math.\ Phys.\  {\bf 7}, 831 (2004)
  [hep-th/0206161].
  
\bibitem{Nekrasov2003} 
  N.~Nekrasov and A.~Okounkov,
  hep-th/0306238.

\bibitem{Aganagic2003} 
  M.~Aganagic, R.~Dijkgraaf, A.~Klemm, M.~Marino and C.~Vafa,
  Commun.\ Math.\ Phys.\  {\bf 261}, 451 (2006)
  [hep-th/0312085].
    
\bibitem{AGGTV} 
  L.~F.~Alday, D.~Gaiotto, S.~Gukov, Y.~Tachikawa and H.~Verlinde,
  JHEP {\bf 1001}, 113 (2010)
  [arXiv:0909.0945 [hep-th]].
  
\bibitem{TongReview} 
  D.~Tong,
  Annals Phys.\  {\bf 324}, 30 (2009)
  [arXiv:0809.5060 [hep-th]].

 
\bibitem{Witten1993} 
  E.~Witten,
  Nucl.\ Phys.\ B {\bf 403}, 159 (1993)
  [hep-th/9301042].


\bibitem{Dorey1998} 
  N.~Dorey,
  JHEP {\bf 9811}, 005 (1998)
  [hep-th/9806056].

\bibitem{Dorey1999} 
  N.~Dorey, T.~J.~Hollowood and D.~Tong,
  JHEP {\bf 9905}, 006 (1999)
  [hep-th/9902134].
   
\bibitem{Shifman2004} 
  M.~Shifman and A.~Yung,
  Phys.\ Rev.\ D {\bf 70}, 045004 (2004)
  [hep-th/0403149].
  
\bibitem{Atiyah1984} 
  M.~F.~Atiyah,
  Commun.\ Math.\ Phys.\  {\bf 93}, 437 (1984).
     

  
  
\bibitem{Awata2010a} 
  H.~Awata and Y.~Yamada,
  Prog.\ Theor.\ Phys.\  {\bf 124}, 227 (2010)
  [arXiv:1004.5122 [hep-th]].

    
\bibitem{HIV} 
  T.~J.~Hollowood, A.~Iqbal and C.~Vafa,
  JHEP {\bf 0803}, 069 (2008)
  [hep-th/0310272].
  
  
\bibitem{Kozcaz2010} 
  C.~Kozcaz, S.~Pasquetti and N.~Wyllard,
  JHEP {\bf 1008}, 042 (2010)
  [arXiv:1004.2025 [hep-th]].

\bibitem{Iqbal2004} 
  A.~Iqbal and A.~-K.~Kashani-Poor,
  Adv.\ Theor.\ Math.\ Phys.\  {\bf 10}, 317 (2006)
  [hep-th/0410174].
  
\bibitem{Gomis2007} 
  J.~Gomis and T.~Okuda,
  JHEP {\bf 0707}, 005 (2007)
  [arXiv:0704.3080 [hep-th]].
  
\bibitem{Kaneko}
J.~Kaneko,
SIAM, J. Math. Anal. {\bf 4} (1993) 1086.

  
\bibitem{OVfactor} 
  H.~Ooguri and C.~Vafa,
  Nucl.\ Phys.\ B {\bf 577}, 419 (2000)
  [hep-th/9912123].

\bibitem{AMV2002} 
  M.~Aganagic, M.~Marino and C.~Vafa,
  Commun.\ Math.\ Phys.\  {\bf 247}, 467 (2004)
  [hep-th/0206164].
    
\bibitem{Bonelli2011} 
  G.~Bonelli, A.~Tanzini and J.~Zhao,
  JHEP {\bf 1206}, 178 (2012)
  [arXiv:1102.0184 [hep-th]].
  
\bibitem{Shadchin2006} 
  S.~Shadchin,
  JHEP {\bf 0708}, 052 (2007)
  [hep-th/0611278].
  
  
\bibitem{Yoshida} 
  Y.~Yoshida,
  arXiv:1101.0872 [hep-th].

 
\bibitem{macdonald}{I.~G.~Macdonald: Symmetric functions and Hall polynomials (Second editions, Oxford Mathematical Monographs)}

\bibitem{Benini2012} 
  F.~Benini and S.~Cremonesi,
  arXiv:1206.2356 [hep-th].

\bibitem{Doroud2012} 
  N.~Doroud, J.~Gomis, B.~Le Floch and S.~Lee,
  arXiv:1206.2606 [hep-th].

\bibitem{Gomis2012} 
  J.~Gomis and S.~Lee,
  arXiv:1210.6022 [hep-th].
  
\bibitem{Chen2012A} 
  H.~-Y.~Chen, T.~J.~Hollowood and P.~Zhao,
  JHEP {\bf 1207}, 139 (2012)
  [arXiv:1205.4230 [hep-th]].

\bibitem{AGT} 
  L.~F.~Alday, D.~Gaiotto and Y.~Tachikawa,
  Lett.\ Math.\ Phys.\  {\bf 91}, 167 (2010)
  [arXiv:0906.3219 [hep-th]].

\bibitem{AGGTV} 
  L.~F.~Alday, D.~Gaiotto, S.~Gukov, Y.~Tachikawa and H.~Verlinde,
  JHEP {\bf 1001}, 113 (2010)
  [arXiv:0909.0945 [hep-th]].
  
\bibitem{AT2010} 
  L.~F.~Alday and Y.~Tachikawa,
  Lett.\ Math.\ Phys.\  {\bf 94}, 87 (2010)
  [arXiv:1005.4469 [hep-th]].

\bibitem{Feigin1994} 
  B.~Feigin, E.~Frenkel and N.~Reshetikhin,
  Commun.\ Math.\ Phys.\  {\bf 166}, 27 (1994)
  [hep-th/9402022].
  
\bibitem{Teschner2010} 
  J.~Teschner,
  Adv.\ Theor.\ Math.\ Phys.\  {\bf 15}, 471 (2011)
  [arXiv:1005.2846 [hep-th]].

\bibitem{Mironov2012A} 
  A.~Mironov, A.~Morozov, Y.~Zenkevich and A.~Zotov,
  Pisma Zh.\ Eksp.\ Teor.\ Fiz.\  {\bf 97}, 49 (2013)
  [arXiv:1204.0913 [hep-th]].
  
\bibitem{Mironov2012B} 
  A.~Mironov, A.~Morozov, B.~Runov, Y.~Zenkevich and A.~Zotov,
  Letters in Mathematical Physics: Volume 10 {\bf 3}, , Page 299 (2013)
  [Lett.\ Math.\ Phys.\  {\bf 103}, 299 (2013)]
  [arXiv:1206.6349 [hep-th]].

\bibitem{BCGK} 
  K.~Bulycheva, H.~-Y.~Chen, A.~Gorsky and P.~Koroteev,
  JHEP {\bf 1210}, 116 (2012)
  [arXiv:1207.0460 [hep-th]].
  
\bibitem{Dimofte2011A} 
  T.~Dimofte, D.~Gaiotto and S.~Gukov,
  arXiv:1108.4389 [hep-th].
  
\bibitem{Dimofte2011B} 
  T.~Dimofte, D.~Gaiotto and S.~Gukov,
  arXiv:1112.5179 [hep-th].
  
\bibitem{Terashima2011} 
  Y.~Terashima and M.~Yamazaki,
  JHEP {\bf 1108}, 135 (2011)
  [arXiv:1103.5748 [hep-th]].


\bibitem{Pasquetti2011} 
  S.~Pasquetti,
  JHEP {\bf 1204}, 120 (2012)
  [arXiv:1111.6905 [hep-th]].
  
\bibitem{Beem2012} 
  C.~Beem, T.~Dimofte and S.~Pasquetti,
  arXiv:1211.1986 [hep-th].
  
  \end{thebibliography}

\end{document}